\documentclass[preprint,sort&compress,12pt]{elsarticle}

\usepackage{graphicx}

\usepackage{amssymb,amsmath}
\usepackage{bm}

\newcommand{\Slash}[1]{\ooalign{\hfil/\hfil\crcr$#1$}}
\newcommand{\bra}[1]{\langle \, #1 \, |}
\newcommand{\ket}[1]{| \, #1 \, \rangle}

\newcommand{\Tcc}{{\rm T}_{\rm cc}}
\newcommand{\three}{[\bm{\bar{3}},{}^{3}{\rm S}_{1}]}
\newcommand{\six}{[\bm{6},{}^{1}{\rm S}_{0}]}

\usepackage[normalem]{ulem}  
\usepackage{color} 

\renewcommand\sout{\bgroup \color{blue} \ULdepth=-.5ex \ULset}

\allowdisplaybreaks[1]

\journal{Nuclear Physics A}

\begin{document}

\begin{frontmatter}



\title{Spectroscopy and production of doubly charmed tetraquarks}


\author[a]{Tetsuo Hyodo}
\ead{hyodo@yukata.kyoto-u.ac.jp}
\author[b]{Yan-Rui Liu}
\author[c,d]{Makoto Oka}
\author[c]{Shigehiro Yasui}
\ead{yasuis@th.phys.titech.ac.jp}

\address[a]{Yukawa Institute for Theoretical Physics, Kyoto University, Kyoto 606-8502, Japan}

\address[b]{School of Physics, Shandong University, Jinan, 250100, P. R. China}

\address[c]{Department of Physics, Tokyo Institute of Technology,
Tokyo 152-8551, Japan}

\address[d]{Advanced Science Research Center, Japan Atomic Energy Agency, Tokai, Ibaraki, 319-1195, Japan}

\begin{abstract}
 We discuss the production of the exotic doubly-charmed tetraquark mesons ${\rm T}_{{\rm c}{\rm c}}({\rm c}{\rm c}\bar{\rm u}\bar{\rm d})$ from electron-positron collisions. ${\rm T}_{{\rm c}{\rm c}}({\rm c}{\rm c}\bar{\rm u}\bar{\rm d})$ is a compact exotic hadron whose binding energy is provided by the diquark correlations. We evaluate the production cross section in the electron-positron collisions using the nonrelativistic QCD framework, and investigate the difference of the production cross section for different color configurations in two charm quarks (antitriplet and sextet), 
where the mixing of different configurations
is suppressed in the heavy quark limit. The total cross section is estimated by modeling the nonperturbative fragmentation process of the charm quark pair into the tetraquark by the wavefunction of the tetraquark and the probability of picking up the light quarks. We find that the internal color configurations are distinguishable by the qualitative features of the differential cross sections which are independent of the nonperturbative matrix elements.
\end{abstract}


\end{frontmatter}



\section{Introduction}

Recent developments of the high energy experiments (Belle, Babar, BESIII, LHCb, \textit{etc.}) have shown the existence of many exotic configuration of hadrons in the heavy quark sector~\cite{Brambilla:2004wf,Swanson:2006st,Voloshin:2007dx,Nielsen:2009uh,Brambilla:2010cs,Esposito:2014rxa,Chen:2016qju,Hosaka:2016pey,Lebed:2016hpi,Ali:2017jda}. To explain the nonconventional properties of the exotic hadrons, various internal structures have been proposed, such as multiquarks, hadronic molecules, gluon hybrids, and so on. Although there is no fundamental principle to forbid the exotic states in QCD, experimentally observed spectrum of hadrons is almost occupied by the conventional configurations~\cite{Olive:2016xmw}.
To understand the rich nonperturbative dynamics of the strong interaction in the heavy hadron spectroscopy, it is important to study how the exotic hadrons are formed and how they are possibly produced in experiments.

There is a candidate of manifestly exotic hadrons in the heavy sector, the tetraquark $\Tcc$ state which contains two charm quarks and two light antiquarks $({\rm cc}\bar{\rm u}\bar{\rm d})$. The isospin and spin-parity quantum numbers are considered to be $I(J^{P})=0(1^{+})$, based on the color-magnetic interaction. Because of the spin-parity, the lowest energy two-meson channel to which $\Tcc$ can couple is the $s$-wave ${\rm D}{\rm D}^{*}$ pair. Because it requires at least four valence quarks, $\Tcc$ is classified as the manifestly exotic hadrons, which contain the pentaquark $\Theta^{+}({\rm uudd}\bar{\rm s})$~\cite{Nakano:2003qx,Nakano:2008ee}. Now, several manifestly exotic hadrons with heavy flavor have been observed, e.g., the charged charmonium-like states ${\rm Z}_{\rm c}$ (${\rm c}\bar{\rm c}{\rm u}\bar{\rm d}$)~\cite{Choi:2007wga,Ablikim:2013mio} and bottomonium-like states ${\rm Z}_{\rm b}({\rm b}\bar{\rm b}{\rm u}\bar{\rm d})$~\cite{Belle:2011aa} (see the reviews~\cite{Swanson:2006st,Brambilla:2010cs} for more information) as well as the new candidate of pentaquark ${\rm P}_{\rm c}$ (${\rm c}\bar{\rm c}{\rm u}{\rm u}{\rm d}$) recently reported by LHCb~\cite{Aaij:2015tga}. 
The existence of $\Tcc$ is theoretically predicted and this tetraquark is still a promising state to be found.
The $\Tcc$ state was originally proposed in the context of the constituent quark models~\cite{Zouzou:1986qh,Lipkin:1986dw} (see also Ref.~\cite{Ader:1981db}). The investigation is further pursued in quark models~\cite{Heller:1986bt,Carlson:1987hh,SilvestreBrac:1993ss,SilvestreBrac:1993ry,Semay:1994ht,Pepin:1996id,SchaffnerBielich:1998ci,Janc:2004qn,Zhang:2007mu}, including  dynamical four-body calculations~\cite{Brink:1998as,Barnea:2006sd,Vijande:2007ix,Vijande:2007fc,Vijande:2007rf,Vijande:2009kj,Vijande:2009zs,Yang:2009zzp}. The role of the strong diquark correlation is emphasized in Refs.~\cite{Ebert:2007rn,Lee:2007tn,Lee:2009rt}. Another approach is based on the hadronic molecule picture, by constructing the potential between ${\rm D}$ and ${\rm D}^{*}$~\cite{Manohar:1992nd,Tornqvist:1993ng,Ding:2009vj,Molina:2010tx,Ohkoda:2011vj,Carames:2011zz}. The ${\rm D}{\rm D}^{*}$ potential is recently studied in lattice QCD~\cite{Ikeda:2013vwa}, and was found to be attractive. The properties of $\Tcc$ are also discussed in the context of the heavy quark symmetry argument~\cite{Cohen:2006jg} and of the QCD sum rules~\cite{Navarra:2007yw,Du:2012wp}.

One of the unique features of the multiquark hadrons is the color configurations. In ordinary mesons and baryons, the quark-quark (quark-antiquark) correlation is limited in the color anti-triplet $\bm{\bar{3}}$ (singlet $\bm{1}$) representation. The color sextet $\bm{6}$ quark-quark correlation (color octet $\bm{8}$ quark-antiquark correlation) is only allowed in the totally color-singlet system with more than three quarks. Thus, the tetraquark is the simplest system in which we can investigate the exotic color correlations. In this paper, we first summarize the spectroscopy of the $\Tcc$ states in the constituent quark model picture with different color configurations. We also discuss the spatial correlations of quarks inside $\Tcc$ with the quark model.

Another purpose of this
paper is to study the production cross section for the orientation of experimental searches. So far, the production yield of heavy exotics including ${\rm T}_{{\rm c}{\rm c}}$ has been estimated for the heavy ion collisions~\cite{Cho:2010db,Cho:2011ew,Cho:2017dcy}. Here we focus on the electron-positron collisions. The double charm production such as ${\rm e}^{+}{\rm e}^{-}\to J/\psi +({\rm c}\bar{\rm c})$ with (${\rm c}\bar{\rm c}$) being any states with a charm quark and a charm antiquark (inclusive process) has been already observed experimentally~\cite{Abe:2002rb,Abe:2004ww,Aubert:2005tj,Pakhlov:2009nj}.
By recombining the charm quarks in the final state of this process, we obtain the production of the doubly charmed hadrons.
In fact, Refs.~\cite{Ma:2003zk,Berezhnoy:2003hz,Jiang:2012jt} have theoretically discussed the production of doubly heavy baryon $\Xi_{cc}$,
which is recently observed in proton-proton collisions by the LHCb collaboration~\cite{Aaij:2017ueg}. 
The production of $\Tcc$ is thus naturally expected as a process ${\rm e}^{+}{\rm e}^{-}\to {\rm T}_{{\rm c}{\rm c}} +(\bar{\rm c}\bar{\rm c})$, 
in the same manner~\cite{Hyodo:2012pm}.
In Ref.~\cite{Hyodo:2012pm}, to estimate the production cross sections, we utilize the nonrelativistic QCD (NRQCD) framework~\cite{Bodwin:1994jh,Petrelli:1997ge}, which is an effective field theory of QCD with the velocity expansion and the factorization of the production processes. 
However, the number of the production of light quarks was not considered in the previous work \cite{Hyodo:2012pm}, and hence we obtained the conclusion that the production cross section of ${\rm T}_{{\rm c}{\rm c}}$ is the same as that of $\Xi_{{\rm c}{\rm c}}$. This is clearly not 
realistic,
because the production cross section of ${\rm T}_{{\rm c}{\rm c}}$ should be smaller than that of $\Xi_{{\rm c}{\rm c}}$ due to the different number of light quarks. In the present work, we consider the factor of the light quark production and estimate the production cross section of ${\rm T}_{{\rm c}{\rm c}}$ in more realistic way.

The paper is organized as follows. In section~\ref{sec:spectrum}, we discuss the mass spectrum of the $\Tcc$ states with various color/flavor configurations based on the diquark picture, and in section~\ref{sec:wavefunction} we calculate the wave functions of $\Tcc$ in the constituent quark model.  In section~\ref{sec:production}, we explain the NRQCD framework for the production of $\Tcc$ in the electron-positron collisions. In section~\ref{sec:numerical}, we present the numerical results of the production cross sections. The last section is devoted to the summary. Based on the analysis in the previous paper Ref.~\cite{Hyodo:2012pm}, we further discuss the decay mode of $\Tcc$, the fragmentation process, and the energy and angular dependence of the cross sections, together with a detailed account of the formulation.

\section{Spectrum and decay of tetraquark $\Tcc$}\label{sec:spectrum}

\subsection{Quark correlation and diquark picture}

The mass of $\Tcc$ has been studied theoretically in quark models by many authors \cite{Ader:1981db,Zouzou:1986qh,Lipkin:1986dw,Heller:1986bt,Carlson:1987hh,SilvestreBrac:1993ss,SilvestreBrac:1993ry,Semay:1994ht,Pepin:1996id,SchaffnerBielich:1998ci,Brink:1998as,Janc:2004qn,Barnea:2006sd,Ebert:2007rn,Lee:2007tn,Vijande:2007ix,Vijande:2007fc,Vijande:2007rf,Zhang:2007mu,Lee:2009rt,Vijande:2009kj,Vijande:2009zs,Yang:2009zzp}. Their results can be essentially understood by the strong diquark correlation in the constituent quark picture~\cite{Lee:2007tn,Lee:2009rt}. To appreciate this, we consider the interaction Hamiltonian through the color-spin interaction~\cite{Jaffe:2004ph};
\begin{align}
    H_{\rm int} 
    = \sum_{i<j} \frac{C_{\rm H}}{m_i m_j} 
    \left(-\frac{3}{8}\right)\vec{\lambda}_{i} \!\cdot\! \vec{\lambda}_{j} \, 
    \vec{s}_{i} \!\cdot\! \vec{s}_{j}.
    \label{eq:H_int}
\end{align}
where $\vec{\lambda}_{i}$ is the Gell-Mann 
matrix
for color SU(3) acting on the quark $i$, and $\vec{s}_{i}=\vec{\sigma}_{i}/2$ is the spin operator with the Pauli matrix $\vec{\sigma}_{i}$ for the quark $i$. We introduce the factor $-3/8$ for the normalization of the pair of quarks in color antitriplet channel; $\langle \bm{\bar{3}} |(-3/8) \vec{\lambda}_{i} \!\cdot\! \vec{\lambda}_{j}|\bm{\bar{3}}\rangle = 1$. The matrix elements of the operator $(-3/8)\vec{\lambda}_{i} \cdot \vec{\lambda}_{j} \, \vec{s}_{i} \cdot \vec{s}_{j}$ for quark pairs in different color-spin channels are summarized in Table~\ref{table:color-spin}. The interaction is specified by the coupling strength $C_{\rm H}$ and the mass of the quark $m_{i}$. In principle, the coupling strength $C_{\rm H}$ reflects the spatial correlation of the quark pair. For instance, with the delta function type interaction, $C_{\rm H}$ is proportional to $\langle \delta(r) \rangle$ with the relative distance of the quark pair $r$. The expectation value is dynamically determined by the wavefunction of each hadron. Here we take a simpler approach~\cite{Lee:2007tn,Lee:2009rt} where $C_{\rm H}$ is taken to be a constant and adjusted to reproduce the mass splittings of known hadrons. Fixing the quark masses as $m_{\rm u}=m_{\rm d}=300$ MeV and $m_{\rm c}=1500$ MeV, we determine the constant $C_{\rm H}=C_{\rm qq}$ for the quark-quark pair by the mass splitting of baryons, $M_{\Delta}-M_{\rm N}$, $M_{\Sigma}-M_{\Lambda}$, $M_{\Sigma_{\rm c}}-M_{\Lambda_{\rm c}}$, and $M_{\Sigma_{\rm b}}-M_{\Lambda_{\rm b}}$. In the same way, the constant $C_{\rm H}=C_{\rm q\bar{q}}$ for the quark-antiquark pair is determined by $M_{\rho}-M_{\pi}$, $M_{{\rm K}^{\ast}}-M_{\rm K}$, $M_{{\rm D}^{\ast}}-M_{\rm D}$, and $M_{{\rm B}^{\ast}}-M_{\rm B}$. We obtain $C_{\rm qq}=(193\text{ MeV})m_{\rm u}^{2}$ and $C_{\rm q\bar{q}}=(318\text{ MeV}) m_{\rm u}^{2}$ which reproduce well the observed mass splittings~\cite{Lee:2009rt}.

In the above determination of $C_{\rm qq}$ and $C_{\rm q\bar{q}}$, the correlation between two charm quarks is not considered. The coupling strength $C_{\rm H}$ for heavy-heavy (cc) quark pair should be modified from those of light-light and heavy-light quark pairs, when the spatial structure is taken into account. Because the kinetic energy is suppressed in the heavy system, the wave function of the heavy-heavy quark pair is spatially shrunk so that the the expectation value $\langle \delta(r) \rangle$ increases. This effectively enhances the coupling strength. In fact, the mass splitting of charmonia ($J/\psi$ and $\eta_{c}$) leads to $C_{\rm c\bar{c}}=(59\text{ MeV})m_{\rm c}^2=(1475\text{ MeV})m_{\rm u}^2$ which is much larger than $C_{\rm qq}$. Because $C_{\rm qq}/C_{\rm q\bar{q}}\sim 2/3$, we determine $C_{\rm cc}=2C_{\rm c\bar{c}}/3=(39\text{ MeV})m_{\rm c}^2=(975\text{ MeV})m_{\rm u}^2$. In practice, the enhancement of the coupling strength in the charm sector does not cause major difference~\cite{Lee:2007tn,Lee:2009rt}, because the heavy-heavy interaction is in any case suppressed by $1/m_{\rm c}^{2}$ factor.

\begin{table}[tbp]
\caption{The matrix elements of the operator $(-3/8)\vec{\lambda}_{i} \!\cdot\! \vec{\lambda}_{j} \, \vec{s}_{i} \!\cdot\! \vec{s}_{j}$ for quark pairs in spin $s=0$, 1 and color $\bm{\bar{3}}$, $\bm{6}$, $\bm{1}$, $\bm{8}$ configurations.}
\begin{center}
\renewcommand{\arraystretch}{1.2}
\begin{tabular}{c|rrrr}
\hline
\hline
  & \multicolumn{1}{c}{$\bm{\bar{3}}$} & \multicolumn{1}{c}{$\bm{6}$} 
  & \multicolumn{1}{c}{$\bm{1}$} & \multicolumn{1}{c}{$\bm{8}$} \\
 \hline
 $s=0$ & $-3/4$ & $3/8$ & $-3/2$ & $3/16$ \\
 \hline
 $s=1$ & $1/4$ & $-1/8$ & $1/2$ & $-1/16$ \\
 \hline
 \hline
\end{tabular}
\end{center}
\label{table:color-spin}
\end{table}%

Many properties of hadron masses are well understood by the diquark picture. To focus on the inter-quark correlations, we regard a pair of quarks as one single cluster called diquark \cite{Lichtenberg:1982jp}.\footnote{We note that the diquark is not necessarily a spatially compact object; the important fact is that there is a correlation of the specific quark pair.} We define the mass of the diquark by the sum of the quark masses and the interaction energy. The latter is calculated from Eq.~\eqref{eq:H_int}. For example, the ud diquark in $\Lambda_{\rm c}$ is combined into color $\bm{\bar{3}}$, spin $^{1}{\rm S}_{0}$ in isospin $I=0$. Because of the attractive quark-quark interaction, the mass of this diquark (so-called ``good" diquark) is about 145 MeV below $2m_{\rm u}$~\cite{Lee:2007tn,Lee:2009rt}.\footnote{The attraction in this channel is a source to form a quark-quark pairing (Cooper pairing) in the color superconductivity in quark matter at high density~\cite{Alford:1998zt,Rapp:1997zu,Alford:2007xm}.} On the other hand, the mass of the ud diquark contained in $\Sigma_{\rm c}$ and $\Sigma_{\rm c}^{\ast}$, with color $\bm{\bar{3}}$, but spin $^{3}{\rm S}_{1}$ in isospin $I=1$ (so-called ``bad" diquark), is about 48 MeV above $2m_{\rm u}$ due to the inter-quark repulsion. Thus we confirm that the mass splitting 
of
several hadrons are well reproduced in the diquark picture.
 This indicates that the diquark picture can be 
used as 
 a good estimation of the masses of exotic heavy hadrons also.

\subsection{Mass spectrum of $\Tcc$ in diquark picture}

We now turn to the study of mass of the exotic tetraquark $\Tcc$ in the diquark picture. The quark content of $\Tcc$ is ${\rm cc}\bar{\rm u}\bar{\rm d}$, which is described by the system of cc diquark plus $\bar{\rm u}\bar{\rm d}$ diquark. We consider the $s$-wave state as the lowest energy configuration. From Table~\ref{table:color-spin}, the most attractive qq configuration is in color $\bm{\bar{3}}$ and spin $^{1}{\rm S}_{0}$. We thus consider the ground state of $\Tcc$ consists of a ``good'' $\bar{\rm u}\bar{\rm d}$ diquark with color $\bm{3}$ and spin $^{1}{\rm S}_{0}$. The antisymmetrization of the quarks leads to the isospin $I=0$. To form the color singlet, the cc diquark should be combined into color $\bm{\bar{3}}$. Because there is no flavor degrees of freedom, the spin of the cc diquark should be $^{3}{\rm S}_{1}$. By combining two diquarks in relative $s$ wave, the total quantum number of $\Tcc$ is given by $I(J^{P})=0(1^{+})$. We note that the cc diquark induces a repulsive interaction, but this is suppressed by the $1/m_{\rm c}^{2}$ factor in the interaction Hamiltonian $H_{\rm int}$ in Eq.~\eqref{eq:H_int}. The dominant contribution of the color-spin interaction is the attraction in the $\bar{\mathrm{u}}\bar{\mathrm{d}}$ diquark.

To investigate the stability of $\Tcc$ in the diquark model, we compare its mass with the lowest two-meson threshold~\cite{Lee:2007tn,Lee:2009rt}. For $I(J^{P})=0(1^{+})$ channel, this is the DD$^{\ast}$ threshold. The binding energy from the DD$^{\ast}$ threshold is estimated as
\begin{align}
    {\rm B.E.}
    &=  \left( -\frac{3}{2} \frac{C_{\rm q\bar{q}}}{m_{\rm u}m_{\rm c}} 
    + \frac{1}{2} \frac{C_{\rm q\bar{q}}}{m_{\rm u}m_{\rm c}} \right)
    -\left(- \frac{3}{4} \frac{C_{\rm qq}}{m_{\rm u}^2} 
    + \frac{1}{4} \frac{C_{\rm cc}}{m_{\rm c}^2} \right)
    \simeq 71\hspace{0.5em} {\rm MeV},
    \label{eq:binding}
\end{align}
where the first term is the interaction energy of the color-spin interaction for ${\rm c}\bar{\rm q}$ pairs in the D and D$^{\ast}$ mesons and the second term is that in $\Tcc$. We note that the interaction energy of the heavy-light pairs are cancelled out in $\Tcc$. Because the binding energy is positive, we find that this state is stable against the strong decay. The obtained binding energy is consistent with the results of the fully dynamical calculation in the constituent quark model~\cite{Brink:1998as,Barnea:2006sd,Vijande:2007ix,Vijande:2007fc,Vijande:2007rf,Vijande:2009kj,Vijande:2009zs,Yang:2009zzp} and that in the QCD sum rule~\cite{Navarra:2007yw}.\footnote{We however note that the QCD sum rule analysis in Ref.~\cite{Du:2012wp} does not conclude the existence of stable doubly charmed tetraquarks.}

Thus the attraction in the $\bar{\rm u}\bar{\rm d}$ diquark in color $\bm{3}$, spin $^{1}{\rm S}_{0}$ and isospin $I=0$ is essential for the stable T$_{\rm cc}$. In addition to this channel, however, there is also an attraction in another channel, color $\bm{\bar{6}}$, spin $^{3}{\rm S}_{1}$ and isospin $I=0$, in $\bar{\rm u}\bar{\rm d}$ diquark (see Table~\ref{table:color-spin}). This suggests the possibility to form a new state of T$_{\rm cc}$ with cc pair in color $\bm{6}$ and spin $^{1}{\rm S}_{0}$ and $\bar{\rm u}\bar{\rm d}$ pair in color $\bm{\bar{6}}$ and spin $^{3}{\rm S}_{1}$. We note the color configurations of the quark pairs ($\bm{6}$ and $\bm{\bar{6}}$) are orthogonal to the conventional T$_{\rm cc}$ discussed above ($\bm{\bar{3}}$ and $\bm{3}$). Although the strength of the attraction is smaller (factor $-1/8$ in Table~\ref{table:color-spin}) than that of the $\bar{\rm u}\bar{\rm d}$ diquark in color $\bm{3}$ (factor $-3/4$), this new configuration can be a candidate of the first excited state of the conventional $\Tcc$. In the following, we refer to the conventional (new) state as $\Tcc\three$ ($\Tcc\six$) labeled by the quantum numbers of the cc pair. The color-spin configuration of quarks is summarized as follows:
\begin{align}
    \ket{\Tcc\three}
    & \sim 
    \ket{
    [\bar{\rm u}\bar{\rm d}]_{\bm{3},^{1}{\rm S}_{0},I=0}
    [{\rm cc}]_{\bm{\bar{3}},^{3}{\rm S}_{1}}
    } , \\
    \ket{\Tcc\six}
    & \sim 
    \ket{
    [\bar{\rm u}\bar{\rm d}]_{\bm{\bar{6}},^{3}{\rm S}_{1},I=0}
    [{\rm cc}]_{\bm{6},^{1}{\rm S}_{0}}
    } .
\end{align}
For later convenience, we present the recombination factor in color space  
\begin{align}
    \ket{
    [\bar{\rm u}\bar{\rm d}]_{\bm{3}}
    [{\rm c}_{1}{\rm c}_{2}]_{\bm{\bar{3}}}
    }
    & = 
    \frac{1}{\sqrt{3}}\ket{
    [\bar{\rm u}{\rm c}_{1}]_{\bm{1}}
    [\bar{\rm d}{\rm c}_{2}]_{\bm{1}}
    } 
    -\sqrt{\frac{2}{3}}\ket{
    [\bar{\rm u}{\rm c}_{1}]_{\bm{8}}
    [\bar{\rm d}{\rm c}_{2}]_{\bm{8}}
    } 
    \label{eq:basis_change_3_1} \\
    & = 
    -\frac{1}{\sqrt{3}}\ket{
    [\bar{\rm u}{\rm c}_{2}]_{\bm{1}}
    [\bar{\rm d}{\rm c}_{1}]_{\bm{1}}
    } 
    +\sqrt{\frac{2}{3}}\ket{
    [\bar{\rm u}{\rm c}_{2}]_{\bm{8}}
    [\bar{\rm d}{\rm c}_{1}]_{\bm{8}}
    }
    \label{eq:basis_change_3_2} .
\end{align}
This suggests that the fraction of color singlet (octet) $\bar{q}q$ pairs is 1/3 (2/3). The wavefunction of $\Tcc\six$ can also be rearranged as 
\begin{align}
    \ket{
    [\bar{\rm u}\bar{\rm d}]_{\bm{\bar{6}}}
    [{\rm c}_{1}{\rm c}_{2}]_{\bm{6}}
    }
    & = 
    \sqrt{\frac{2}{3}}\ket{
    [\bar{\rm u}{\rm c}_{1}]_{\bm{1}}
    [\bar{\rm d}{\rm c}_{2}]_{\bm{1}}
    } 
    +\frac{1}{\sqrt{3}}\ket{
    [\bar{\rm u}{\rm c}_{1}]_{\bm{8}}
    [\bar{\rm d}{\rm c}_{2}]_{\bm{8}}
    } 
    \label{eq:basis_change_6_1} \\
    & = 
    \sqrt{\frac{2}{3}}\ket{
    [\bar{\rm u}{\rm c}_{2}]_{\bm{1}}
    [\bar{\rm d}{\rm c}_{1}]_{\bm{1}}
    } 
    +\frac{1}{\sqrt{3}}\ket{
    [\bar{\rm u}{\rm c}_{2}]_{\bm{8}}
    [\bar{\rm d}{\rm c}_{1}]_{\bm{8}}
    }
    \label{eq:basis_change_6_2} .
\end{align}

Because 
the total quantum number of $\Tcc\six$ is $I(J^{P})=0(1^{+})$, which is the same with $\Tcc\three$, 
one may wonder
that $\Tcc\six$ can strongly mix with $\Tcc\three$, for instance, through the one gluon exchange. The transition process however needs to flip the spin of the heavy quark, because of the different spin configuration of the cc pair in $\Tcc\six$ from that in $\Tcc\three$. Since the spin-flip amplitude is suppressed by $1/m_{\rm c}$, the mixing probability of $\Tcc\three$ and $\Tcc\six$ should be of the order of $1/m_{\rm c}^2$. In the heavy quark limit, this mixing must vanish, while the splitting between $\Tcc\three$ and $\Tcc\six$ remains as it comes from the mass difference of light diquarks (See Eq.~\eqref{eq:splitting} below). In a fully dynamical four-body quark-model calculation~\cite{Vijande:2009zs}, the ground state of T$_{\rm cc}$ is almost dominated ($\sim 88$ \%) by the component with the cc pair in color $\bm{\bar{3}}$. It is therefore a good approximation to neglect the mixing effect. It should be emphasized that the suppression of the mixing is ensured by the spin structure of the heavy diquark regardless to any model settings.

Based on the suppression of the mixing, we estimate the mass splitting between $\Tcc\three$ and $\Tcc\six$ by the color-spin interaction~\eqref{eq:H_int} as
\begin{align}
    M({\rm T}_{\rm cc}[\bm{6},{}^{1}{\rm S}_{0}])
    -M({\rm T}_{\rm cc}[\bm{\bar{3}},^{3}\!{\rm S}_{1}])
    &=
    \left(- \frac{1}{8} \frac{C_{\rm qq}}{m_{\rm u}^{2}} 
    + \frac{3}{8} \frac{C_{\rm cc}}{m_{\rm c}^{2}}\right)
    - \left( -\frac{3}{4} \frac{C_{\rm qq}}{m_{\rm u}^{2}} 
    + \frac{1}{4} \frac{C_{\rm cc}}{m_{\rm c}^{2}} \right) 
    \nonumber \\
    &\simeq 125 \hspace{0.5em} {\rm MeV}.
    \label{eq:splitting}
\end{align}
Thus, the mass of $\Tcc\six$ is larger than that of $\Tcc\three$ by 125 MeV. In the heavy quark limit, $M({\rm T}_{\rm cc}[\bm{6},{}^{1}{\rm S}_{0}])-M({\rm T}_{\rm cc}[\bm{\bar{3}},^{3}\!{\rm S}_{1}]) \to (5/8) C_{\rm qq}/m_{\rm u}^{2}\simeq 121$ MeV.

%
\begin{figure}
\begin{center}
\vspace*{-3em}
\hspace*{-2em}
\includegraphics*[width=16cm]{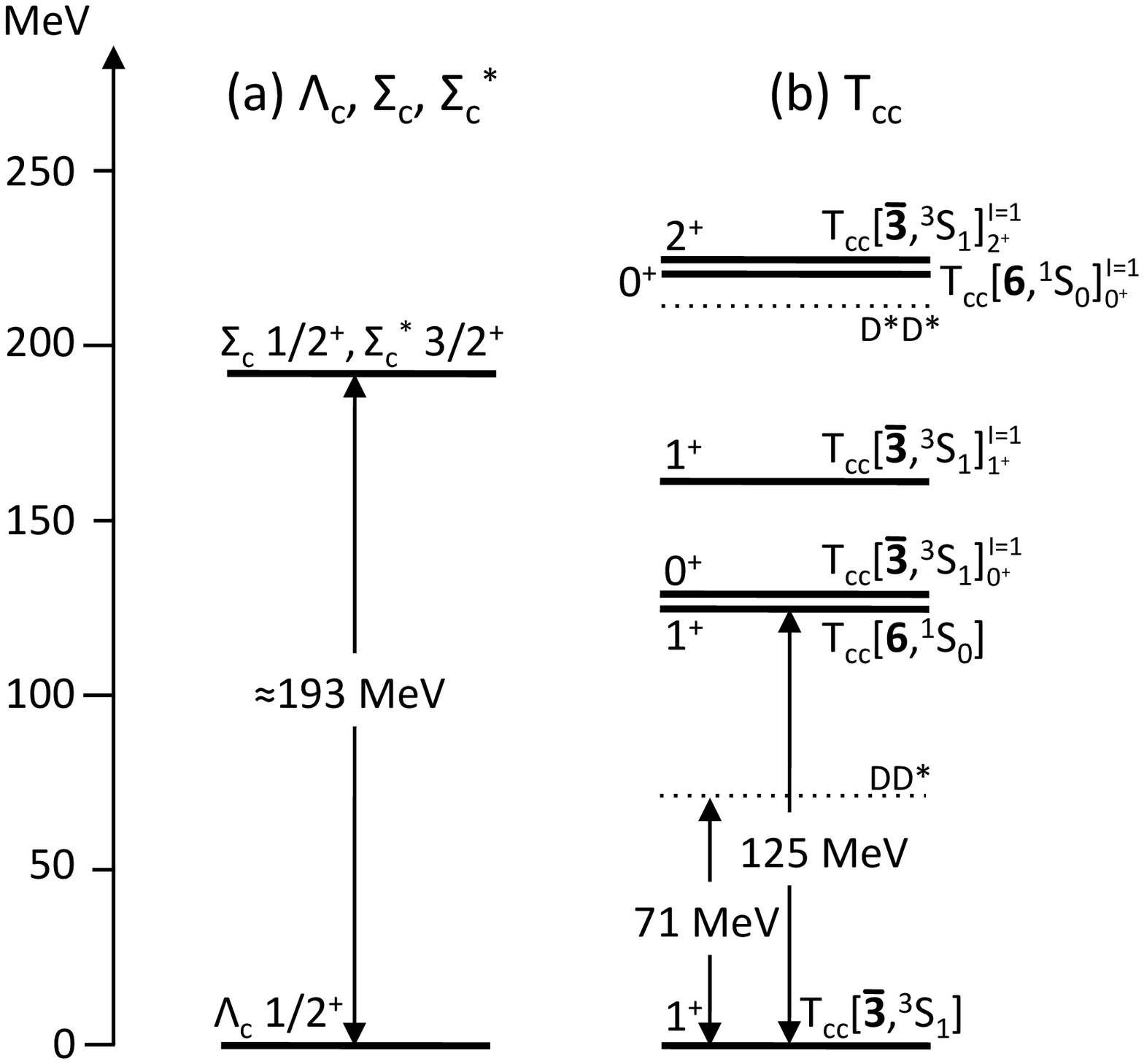}
\vspace*{-3em}
\caption{Mass spectra of (a) $\Lambda_{\mathrm c}$, $\Sigma_{\mathrm c}$ and $\Sigma_{\mathrm c}^{\ast}$ baryons and of (b) $\mathrm{T}_{\mathrm{cc}}$.
The quantum numbers of two charm quarks ($\mathrm{c}\mathrm{c}$)  and two light antiquarks ($\bar{\mathrm q}\bar{\mathrm q}$) are shown.}
\label{fig:Fig_diagram}
\end{center}
\end{figure}
%

Next we consider the stability of $\Tcc\six$ via the strong interaction.\footnote{Possible weak and radiative decays will be discussed in section~\ref{subsec:decay}.} From the isospin conservation, the transition from $\Tcc\six$ into $\Tcc\three$ should emit at least two pions. This is kinematically forbidden by the mass difference in Eq.~\eqref{eq:splitting}. Combining the binding energy of $\Tcc\three$~\eqref{eq:binding} and the mass splitting~\eqref{eq:splitting}, we find that $\Tcc\six$ is located by 54 MeV above the DD$^{\ast}$ threshold (see Fig.~\ref{fig:Fig_diagram}). Thus, $\Tcc\six$ decays into D and D$^{\ast}$ through the fall-apart process in $s$ wave, and hence that the decay width of  $\Tcc\six$ may not be negligibly small. However, the color recombination in the process $[\bar{\rm u}\bar{\rm d}]_{\bm{\bar{6}}}[{\rm cc}]_{\bm{6}}\to [\bar{\rm u}{\rm c}]_{\bm{1}}[\bar{\rm d}{\rm c}]_{\bm{1}}$ can suppress the decay width of $\Tcc\six$.

Let us investigate the other possible excited states with higher masses for $\Tcc$.
For example, we may consider the $I=1$ ud diquarks in ${\rm c}{\rm c}\bar{\rm u}\bar{\rm d}$. This combination leads to $\Tcc$ with $I(J^{P})=1(0^{+})$, $1(1^{+})$, and $1(2^{+})$. Those states are denoted by $\Tcc\three_{0^{+},1^{+},2^{+}}^{I=1}$ and $\Tcc\six_{0^{+}}^{I=1}$.
Unlike the $I=0$ case where no correlation between the heavy quark and the light antiquark exists, now the heavy-light correlation does not vanish. One may calculate its contributions in color and spin spaces separately and then combine the results together. Similar to Eqs. (5) and (6), we may get the recombination factor in spin space
\begin{align}
|[\bar{u}\bar{d}]_{s=1}[c_1c_2]_{s=1}\rangle_{J=0}
&=\frac{\sqrt3}{2}|[\bar{u}c_1]_{s=0}[\bar{d}c_2]_{s=0}\rangle_{J=0}-\frac12|[\bar{u}c_1]_{s=1}[\bar{d}c_2]_{s=1}\rangle_{J=0}\nonumber\\
&=\frac{\sqrt3}{2}|[\bar{u}c_2]_{s=0}[\bar{d}c_1]_{s=0}\rangle_{J=0}-\frac12|[\bar{u}c_2]_{s=1}[\bar{d}c_1]_{s=1}\rangle_{J=0},\\
|[\bar{u}\bar{d}]_{s=1}[c_1c_2]_{s=1}\rangle_{J=1}
&=\frac{1}{\sqrt2}|[\bar{u}c_1]_{s=1}[\bar{d}c_2]_{s=0}\rangle_{J=1}+\frac{1}{\sqrt2}|[\bar{u}c_1]_{s=0}[\bar{d}c_2]_{s=1}\rangle_{J=1}\nonumber\\
&=\frac{1}{\sqrt2}|[\bar{u}c_2]_{s=1}[\bar{d}c_1]_{s=0}\rangle_{J=1}+\frac{1}{\sqrt2}|[\bar{u}c_2]_{s=0}[\bar{d}c_1]_{s=1}\rangle_{J=1},\\
|[\bar{u}\bar{d}]_{s=1}[c_1c_2]_{s=1}\rangle_{J=2}
&=|[\bar{u}c_1]_{s=1}[\bar{d}c_2]_{s=1}\rangle_{J=2}
=|[\bar{u}c_2]_{s=1}[\bar{d}c_1]_{s=1}\rangle_{J=2};\\
|[\bar{u}\bar{d}]_{s=0}[c_1c_2]_{s=0}\rangle_{J=0}
&=\frac{\sqrt3}{2}|[\bar{u}c_1]_{s=1}[\bar{d}c_2]_{s=1}\rangle_{J=0}+\frac12|[\bar{u}c_1]_{s=0}[\bar{d}c_2]_{s=0}\rangle_{J=0}\nonumber\\
&=-\frac{\sqrt3}{2}|[\bar{u}c_2]_{s=1}[\bar{d}c_1]_{s=1}\rangle_{J=0}-\frac12|[\bar{u}c_2]_{s=0}[\bar{d}c_1]_{s=0}\rangle_{J=0}.
\end{align}
Then the mass difference of $\Tcc\three^{I=1}_{0^{+},1^{+},2^{+}}$ from the ground state $\Tcc\three$ is estimated as
\begin{align}
&\quad M(\Tcc\three^{I=1}_{0^+,1^+,2^+})-M(\Tcc\three) \nonumber\\
&=\Big(\frac14\frac{C_{qq}}{m_u^2}+\frac14\frac{C_{cc}}{m_c^2}+x\frac{C_{q\bar{q}}}{m_um_c}\Big)
-\Big(-\frac34\frac{C_{qq}}{m_u^2}+\frac14\frac{C_{cc}}{m_c^2}\Big)\nonumber\\
&=\frac{C_{qq}}{m_u^2}+x\frac{C_{q\bar{q}}}{m_um_c}\simeq (193+64x) \text{ MeV}
\end{align}
where $x=-1$, $-\frac12$, and $\frac12$ for $0^+$, $1^+$, and $2^+$, respectively. Explicitly, the corresponding values are 129 MeV, 161 MeV, and 225 MeV. The mass difference of $\Tcc\six^{I=1}_{0^+}$ from $\Tcc\six$ is
\begin{align}
M(\Tcc\six^{I=1}_{0^+})-M(\Tcc\six)
&=\Big(\frac38\frac{C_{qq}}{m_u^2}+\frac38\frac{C_{cc}}{m_c^2}\Big)
-\Big(-\frac18\frac{C_{qq}}{m_u^2}+\frac38\frac{C_{cc}}{m_c^2}\Big)\nonumber\\
&=\frac12\frac{C_{qq}}{m_u^2}\simeq 96.5 \text{ MeV}.
\end{align}
Here, we have used the matrix elements in Table~\ref{table:color-spin}. 
One may use this method to check that there is no heavy-light correlation in the $I=0$ states
\begin{align}
|[\bar{u}\bar{d}]_{s=0}[c_1c_2]_{s=1}\rangle_{J=1}
&=\frac{1}{\sqrt2}|[\bar{u}c_1]_{s=1}[\bar{d}c_2]_{s=1}\rangle_{J=1}-\frac12|[\bar{u}c_1]_{s=1}[\bar{d}c_2]_{s=0}\rangle_{J=1} \nonumber \\
&\quad +\frac12|[\bar{u}c_1]_{s=0}[\bar{d}c_2]_{s=1}\rangle_{J=1}\\
&=\frac{1}{\sqrt2}|[\bar{u}c_2]_{s=1}[\bar{d}c_1]_{s=1}\rangle_{J=1}-\frac12|[\bar{u}c_2]_{s=1}[\bar{d}c_1]_{s=0}\rangle_{J=1} \nonumber\\
&\quad +\frac12|[\bar{u}c_2]_{s=0}[\bar{d}c_1]_{s=1}\rangle_{J=1},\\
|[\bar{u}\bar{d}]_{s=1}[c_1c_2]_{s=0}\rangle_{J=1}
&=\frac{1}{\sqrt2}|[\bar{u}c_1]_{s=1}[\bar{d}c_2]_{s=1}\rangle_{J=1}+\frac12|[\bar{u}c_1]_{s=1}[\bar{d}c_2]_{s=0}\rangle_{J=1} \nonumber \\
&\quad -\frac12|[\bar{u}c_1]_{s=0}[\bar{d}c_2]_{s=1}\rangle_{J=1}\\
&=-\frac{1}{\sqrt2}|[\bar{u}c_2]_{s=1}[\bar{d}c_1]_{s=1}\rangle_{J=1}-\frac12|[\bar{u}c_2]_{s=1}[\bar{d}c_1]_{s=0}\rangle_{J=1} \nonumber\\
&\quad +\frac12|[\bar{u}c_2]_{s=0}[\bar{d}c_1]_{s=1}\rangle_{J=1}.
\end{align}
As a result, we obtain the mass spectrum of doubly charmed tetraquarks shown in Fig.~\ref{fig:Fig_diagram} (b). It is instructive to compare this spectrum with the corresponding baryon spectrum in Fig.~\ref{fig:Fig_diagram} (a). The tetraquarks with color $\bm{\bar{3}}$ correlation have their counterpart in the baryon spectrum, while the color $\bm{6}$ states are only allowed in $\Tcc$. The exotic color $\bm{6}$ correlation is a unique feature of the tetraquark state, which is not accessible in ordinary hadrons.

Because the $I=1$ states are much heavier than the two-body decay threshold, the effect of the strong decay may be significant. For instance, the lowest thresholds in $I=1$ sector are $\mathrm{D}\mathrm{D}$ in $s$ wave and $d$ wave for $I(J^P)=1(0^+)$ and $1(2^+)$, respectively, and $\mathrm{D}\mathrm{D}^{\ast}$ in $s$ wave for $1(1^{+})$ (see e.g. Table~I in Ref.~\cite{Ohkoda:2012hv}). In the following, therefore, we discuss the structure and production of the $I=0$ states.

In a separate work~\cite{Luo:2017eub}, the mixing effects between states with the same quantum numbers caused by the color-spin interaction are considered. It is found that the mass gap between the two $I(J^{P})=1(0^+)$ states changes from $\sim 100$ MeV to $\sim$ 280 MeV and that between the two $I(J^{P})=0(1^+)$ states from $\sim$130 MeV to $\sim$190 MeV. Alternatively, the mass shift for the former states is around 90 MeV while that for the latter states is around 30 MeV. This observation means that the state mixing is small for the $I(J^{P})=0(1^+)$ tetraquarks (as expected) and is large for the $I(J^{P})=1(0^+)$ tetraquarks. As a result, the decay properties for the $I(J^{P})=1(0^+)$ states may be affected. In the following discussions, we mainly focus on the isoscalar tetraquarks $\Tcc\three$ and $\Tcc\six$. When we consider the $\Tcc$ production, the $I(J^{P})=1(0^+)$ states will also be discussed.

\subsection{Decay modes of T$_{\rm cc}$}\label{subsec:decay}

For experimental searches, it is important to specify the decay properties. Possible decay modes of $\Tcc\three$ and $\Tcc\six$ are summarized in Table~\ref{table:Tcc_decay}. Because $\Tcc\three$ is stable against the strong decay in our estimation, it can decay only by the weak and electromagnetic interactions. In contrast, $\Tcc\six$ decays via the strong interaction which is the dominant contribution to the total width of $\Tcc\six$. Because $\Tcc\six$ has positive charge, the final state in the strong decay is $\mathrm{D}^{0}\mathrm{D}^{\ast+}$ and/or $\mathrm{D}^{\ast 0}\mathrm{D}^{+}$. The direct decay into $\mathrm{D}\mathrm{D}\pi$ state (with one of the particles being positively charged) is also possible.

To consider the weak decay, we recall the rearrangement of color configuration of $\Tcc$ in Eqs.~\eqref{eq:basis_change_3_1}-\eqref{eq:basis_change_6_2}. The weak decay of $\Tcc$ is caused by the decay of D meson in the $\mathrm{D}\mathrm{D}^{\ast}$ ($\mathrm{D}^{0}\mathrm{D}^{\ast+}$ or $\mathrm{D}^{\ast0}\mathrm{D}^{+}$) component. The final state can be, for instance, ${\rm D}^{*}\bar{{\rm K}}\pi$, ${\rm D} \bar{{\rm K}}\pi\pi$, ${\rm D}^{*}\bar{{\rm K}}\pi\pi\pi$, and so on. Among others, $\mathrm{D}^{\ast+}\mathrm{K}^{-}\pi^{+}$ and $\mathrm{D}^{\ast+}\mathrm{K}^{-}\pi^{+}\pi^{+}\pi^{-}$ are advantageous in experimental searches, because the final states are all charged particles.

The radiative decay of $\Tcc$ occurs through the ${\rm D}^{*0}\to {\rm D}^{0}\gamma$ process, which is about 38 \% of the ${\rm D}^{*0}$ decay \cite{Olive:2016xmw}. According to Eqs.~\eqref{eq:basis_change_3_1}-\eqref{eq:basis_change_6_2}, the fraction of the $\mathrm{D}\mathrm{D}^{\ast}$ component is $1/3$ in $\Tcc\three$ and 2/3 in $\Tcc\six$. This indicates the ratio of the radiative decays of $\Tcc\three$ and $\Tcc\six$ will be 
\begin{eqnarray}
\frac{\mathrm{Br}({\mathrm T}_{\rm cc}[\bm{\bar{3}},{}^{3}{\rm S}_{1}] \rightarrow \mathrm{D}^{+} \mathrm{D}^{0}\gamma)}
{\mathrm{Br}(\mathrm{ T}_{\rm cc}[\bm{6},{}^{1}{\rm S}_{0}] \rightarrow \mathrm{D}^{+} \mathrm{D}^{0}\gamma)} = \frac{1}{2}.
\end{eqnarray}
Thus, the measurement of the radiative decay will provide an interesting method to discriminate the internal color structures of T$_{\rm cc}[\bm{\bar{3}},{}^{3}{\rm S}_{1}]$ and T$_{\rm cc}[\bm{6},{}^{1}{\rm S}_{0}]$, in addition to the information from the mass spectroscopy and the production process, as discussed in section \ref{sec:production}.

\renewcommand{\arraystretch}{1.2}
\begin{table}[btp]
\caption{Possible decay modes of $\Tcc\three$ and $\Tcc\six$. Weak decays with all charged final states are shown.}
\begin{center}
\begin{tabular}{c|c|c}
\hline
 & $\Tcc\three$ & $\Tcc\six$ \\
\hline
strong decay & --- & $\mathrm{D}^{0}\mathrm{D}^{\ast+}$, $\mathrm{D}^{\ast 0}\mathrm{D}^{+}$\\
&& $\mathrm{D}^{0}\mathrm{D}^{0}\pi^{+}$, $\mathrm{D}^{0}\mathrm{D}^{+}\pi^{0}$ \\
\hline
weak decay & $\mathrm{D}^{+}\mathrm{K}^{-}\pi^{+}$, $\mathrm{D}^{\ast+}\mathrm{K}^{-}\pi^{+}$, $\mathrm{D}^{\ast+}\mathrm{K}^{-}\pi^{+}\pi^{+}\pi^{-}$ & same with $\Tcc\three$ \\
\hline
radiative decay & $\mathrm{D}^{+} \mathrm{D}^{0}\gamma$ & same with $\Tcc\three$ \\
\hline
\end{tabular}
\end{center}
\label{table:Tcc_decay}
\end{table}%
\renewcommand{\arraystretch}{1}

\section{Wave function of tetraquark $\Tcc$}
\label{sec:wavefunction}

In this section, we discuss the spatial wave function of tetraquark $\Tcc$ in the non-relativistic constituent quark model. The spatial correlation of the cc pair in $\Tcc$ is important to estimate the total cross section of the inclusive production of $\Tcc$ in the next section. We first determine the model parameters by using the masses of normal charm hadrons, and then apply the model setup to $\Tcc$.

\subsection{Constituent quark model with harmonic oscillator potential}

As the simple model, we consider the quark model with the harmonic oscillator potential as the color confinement potential. The interaction between quarks 1 and 2 is given by
\begin{align}
    V_{12}(r) 
    &=  
    \left( -\frac{3}{16} 
    \vec{\lambda}_{1} \cdot \vec{\lambda}_{2} \right) 
    \frac{k}{2} r_{12}^2,\label{eq:HO_potential}
\end{align}
with the strength constant $k$ and the relative distance between the two particle $r_{12}$. The matrix element of the color factor $\vec{\lambda}_{1} \cdot \vec{\lambda}_{2}$ are summarized in Table~\ref{table:lambda_lambda}. In Eq.~\eqref{eq:HO_potential}, the strength is normalized as unity for the single $\bar{q}q$ channel.
Let us consider the q$\bar{\rm q}$ system with reduced mass $\mu$. By defining $\mu \, \omega^2 = (-3/16)\bra{\bm{1}}\vec{\lambda}_{1} \cdot \vec{\lambda}_{2}\ket{\bm{1}} \, k$,
the Hamiltonian for this system is 
\begin{align}
    H_{\rm{HO}}(\vec{x})
    &=  
    - \frac{1}{2\mu} \frac{\partial^2}{\partial \vec{x}^{\,2}} 
    + \frac{\mu \, \omega^2}{2} \vec{x}^{\,2} ,
    \label{eq:HO_hamiltonian}
\end{align}
where $\vec{x}=(x,y,z)$ is the relative coordinate. The solution of the Schr\"odinger equation is specified by the non-negative integers $\{n\}=\{n_1,n_2,n_3\}$ as 
\begin{align}
    \varphi_{\{n\}}(x,y,z)
    &=  
    C_{\{n\}} e^{-\frac{\mu\omega}{2}(x^2+y^2+z^2)} 
    H_{n_1}(\sqrt{\mu\,\omega}\,x) H_{n_2}(\sqrt{\mu\,\omega}\,y) 
    H_{n_3}(\sqrt{\mu\,\omega}\,z)
\end{align}
with the Hermite polynomials $H_{n}$ and a normalization constant $C_{\{n\}}$. For the ground state with $\{n\}=\{0,0,0\}\equiv \{0\}$, the normalized solution is given by
\begin{align}
    \varphi_{\{0\}}(x,y,z)
    &=  
    \left( \frac{\mu\,\omega}{\pi} \right)^{3/4} 
    e^{-\frac{\mu\omega}{2}(x^2+y^2+z^2)}.
    \label{eq:zero_solution}
\end{align}
The root-mean-squared-radius $\sqrt{\langle \vec{x}^{\,2} \rangle}$ for the ground state is calculated as 
\begin{align}
    \sqrt{\langle \vec{x}^{\,2} \rangle} 
    &=  
    \left[ \int \varphi_{\{0\}}^{\ast}(\vec{x}) \, \vec{x}^{\,2} 
    \varphi_{\{0\}}(\vec{x}) \mathrm{d}^3 \vec{x} \, \right]^{1/2} 
    \nonumber \\
    &=  \sqrt{\frac{3}{2}} \frac{1}{\sqrt{\mu\,\omega}}.
\label{eq:mean_squared_radius}
\end{align}

\begin{table}[btp]
\caption{The expectation values of $\vec{\lambda}_{1} \cdot \vec{\lambda}_{2}$ for $\bar{q}q$ and $qq$ channels.}
\begin{center}
\begin{tabular}{cc cc}
\hline
\multicolumn{2}{c}{$\bar{q}q$} & \multicolumn{2}{c}{$qq$} \\
\hline
${\bf 1}_{\mathrm c}$ & ${\bf 8}_{\mathrm c}$ & $\bar{\bf 3}_{\mathrm c}$ & ${\bf 6}_{\mathrm c}$ \\
\hline
$-16/3$ & 2/3 & $-8/3$ & 4/3 \\
\hline
\end{tabular}
\end{center}
\label{table:lambda_lambda}
\end{table}%

\subsection{Parameter fitting to charmonium mass}
\label{sec:charmonium}

In determination of the model parameters, what is the most important in our discussion is the value of the wave function at the center-of-mass for ${\rm c}{\rm c}$ quark pair.
In the realistic calculation with the Cornel (linear-type) potential by Barnes, Godfrey and Swanson \cite{Barnes:2005pb}, they obtained the values
\begin{eqnarray}
|R_{\eta_{\rm c}}(0)|^2 = (1.39)^2 \hspace{0.5em}\mathrm{GeV}^3, \\
|R_{\rm{J}/\psi}(0)|^2 = (1.10)^2 \hspace{0.5em}\mathrm{GeV}^3,
\end{eqnarray}
for $\eta_{\rm c}$ and $\rm{J}/\psi$, respectively.
We use the averaged value in spin;
\begin{eqnarray}
\frac{|R_{\eta_{\rm c}}(0)|^2 + 3|R_{\rm{J}/\psi}(0)|^2}{4} = (1.18)^2 \hspace{0.5em}\mathrm{GeV}^3.
\label{eq:BGS0}
\end{eqnarray}
In their calculation, the charm quark mass is $m_{\rm c}=1.48$ GeV.

In our simple harmonic oscillator potential, the corresponding value is given as
\begin{eqnarray}
R_{{\rm c}\bar{\rm c}}(0) = \frac{2}{b^{3/2}},
\end{eqnarray}
with $b=1/\sqrt{\mu\,\omega}$ from Eq.~(\ref{eq:zero_solution}).
When we require that $|R_{{\rm c}\bar{\rm c}}(0)|^{2}=(1.18)^2$ GeV$^{3}$ (Eq.~(\ref{eq:BGS0})), we obtain $\omega = 0.699$ GeV and
$k = \mu \,\omega^2 = 0.331 \hspace{0.5em}\mathrm{GeV}^{3}$.
In the followings, we use this value, not only for the $\bar{\mathrm{c}}{\mathrm c}$ channel, but also for the $\mathrm{cc}$ channel.
The potential for the latter is easily obtained by changing the color factor in Eq.~(\ref{eq:HO_potential}).

However, there is some ambiguity in determination of the parameters, $m_{\rm c}$ and $\omega$.
We can use another way to reproduce the mass splitting between the $s$-wave charmonia ($\eta_{\rm c}$, $J/\psi$) and $p$-wave charmonia ($h_{\rm c}$ and $\chi_{{\rm c}J}$ ($J=0,1,2$)), which corresponds to the frequency $\omega$ in the harmonic oscillator potential $V(r)=(\mu \, \omega^2 /2) r^2$.
We choose $\omega=0.295$ GeV with the charm quark mass $m_{\rm c}=1.326$ GeV ($\mu = m_{\rm c}/2$) \cite{Flamm:1982jv}. The corresponding value of the spring constant in the present potential is $k=0.0577$ GeV$^{3}$.
In this case, we obtain $|R_{{\rm c}\bar{\rm c}}(0)|^2=(0.59)^2$ GeV$^{3}$. We will discuss how the results change in the production cross section of $\Tcc$ for different parameter sets in section \ref{sec:production}.


\subsection{$\mathrm{T}_{\mathrm{cc}}$}
\label{sec:Tcc_hamiltonian}

$\mathrm{T}_{\mathrm{cc}}$ is the four-body system with the quark content ${\mathrm c}{\mathrm c}\bar{\mathrm u}\bar{\mathrm d}$.
From the quark-quark interaction in Eq.~(\ref{eq:HO_potential}), the Hamiltonian is given as
\begin{eqnarray}
H(\vec{r}_1,\vec{r}_2,\vec{r}_3,\vec{r}_4) 
\!\!\!&=&\!\!\! K + V \nonumber \\
\!\!\!&=&\!\!\!
\sum_{i=1}^{4} \left(-\frac{1}{2m_{i}} \vec{\nabla}_{i}^{2} \right) + \left(-\frac{3}{16} \right) \frac{k}{2} \sum_{i<j} \vec{\lambda}_{i} \cdot \vec{\lambda}_{j} (\vec{r}_{i}-\vec{r}_{j})^2,
\label{eq:Tcc_hamiltonian}
\end{eqnarray}
for four quarks $i=1, 2, 3, 4$.
This is our basic Hamiltonian for $\Tcc\three$ and $\Tcc\six$.

\subsubsection{$\Tcc\three$}
\label{sec:Tcc3}

We assume that the pair of two quarks (antiquarks) belong to color $\bar{\bf 3}_{\mathrm c}$ (${\bf 3}_{\mathrm c}$).
In the four-body system, however, we cannot single out the color factor $\vec{\lambda}_{i} \cdot \vec{\lambda}_{j}$, because it acts not only to the {\it quark-quark} pair with $\bar{\bf 3}_{\mathrm c}$ but also the {\it quark-antiquark} pair with ${\bf 1}_{\mathrm c}$ or ${\bf 8}_{\mathrm c}$.
Then, we obtain the values of the color factor $\vec{\lambda}_{i} \cdot \vec{\lambda}_{j}$ ($i<j$);
\begin{eqnarray}
\vec{\lambda}_{1} \cdot \vec{\lambda}_{2} &=& -\frac{8}{3}, \nonumber \\
\vec{\lambda}_{1} \cdot \vec{\lambda}_{3} &=& \frac{1}{3} \left( -\frac{16}{3} \right) + \frac{2}{3} \frac{2}{3} = -\frac{4}{3}, \nonumber \\
\vec{\lambda}_{1} \cdot \vec{\lambda}_{4} &=& \frac{1}{3} \left( -\frac{16}{3} \right) + \frac{2}{3} \frac{2}{3} = -\frac{4}{3}, \nonumber \\
\vec{\lambda}_{2} \cdot \vec{\lambda}_{3} &=& \frac{1}{3} \left( -\frac{16}{3} \right) + \frac{2}{3} \frac{2}{3} = -\frac{4}{3}, \nonumber \\
\vec{\lambda}_{2} \cdot \vec{\lambda}_{4} &=& \frac{1}{3} \left( -\frac{16}{3} \right) + \frac{2}{3} \frac{2}{3} = -\frac{4}{3}, \nonumber \\
\vec{\lambda}_{3} \cdot \vec{\lambda}_{4} &=& -\frac{8}{3},
\end{eqnarray}
which are obtained by $\langle {\bf 3}_{12}\bar{\bf 3}_{34} | \vec{\lambda}_{i} \cdot \vec{\lambda}_{j} | {\bf 3}_{12}\bar{\bf 3}_{34} \rangle$ with using Eqs.~(\ref{eq:basis_change_3_1}) and (\ref{eq:basis_change_3_2}).
As a consequence, the potential in Eq.~(\ref{eq:Tcc_hamiltonian}) becomes
\begin{eqnarray}
V &=&
\frac{k}{2}
\left\{ \frac{1}{2} \left( \vec{r}_1-\vec{r}_2 \right)^2
      + \frac{1}{4} \left( \vec{r}_1-\vec{r}_3 \right)^2
      + \frac{1}{4} \left( \vec{r}_1-\vec{r}_4 \right)^2 \right. \nonumber \\
&&\hspace{0.8em} \left.
      + \frac{1}{4} \left( \vec{r}_2-\vec{r}_3 \right)^2
      + \frac{1}{4} \left( \vec{r}_2-\vec{r}_4 \right)^2
      + \frac{1}{2} \left( \vec{r}_3-\vec{r}_4 \right)^2
       \right\}.
\end{eqnarray}

Now, let us introduce the relative coordinates $\vec{\xi}$, $\vec{\eta}$ and $\vec{\zeta}$ and the center-of-mass coordinate $\vec{R}$, which are defined by
\begin{eqnarray}
\left(
\begin{array}{c}
 \vec{\xi} \\
 \vec{\eta} \\
 \vec{\zeta} \\
 \vec{R}
\end{array}
\right)
=
\left(
\renewcommand{\arraystretch}{1.2}
\begin{array}{cccc}
 1 & -1 & 0 & 0 \\
 \frac{m_1}{m_{12}} & \frac{m_2}{m_{12}} & -1 & 0 \\
 \frac{m_1}{m_{123}} & \frac{m_2}{m_{123}} & \frac{m_3}{m_{123}} & -1 \\
 \frac{m_1}{M} & \frac{m_2}{M} & \frac{m_3}{M} & \frac{m_4}{M} \\
\end{array}
\right)
\left(
\begin{array}{c}
 \vec{r}_1 \\
 \vec{r}_2 \\
 \vec{r}_3 \\
 \vec{r}_4
\end{array}
\right), \nonumber \\
\label{eq:relative_4_a}
\end{eqnarray}
with the original coordinate $\vec{r}_{i}$ for the (anti)particle $i$.
Here we define $m_{12}=m_{1}+m_{2}$, $m_{123}=m_{1}+m_{2}+m_{3}$ and $M=m_{1}+m_{2}+m_{3}+m_{4}$ (the center-of-mass).
The kinetic term is then rewritten as
\begin{eqnarray}
K = -\frac{1}{2\mu} \vec{\nabla}_{\xi}^{2} - \frac{1}{2\nu} \vec{\nabla}_{\eta}^{2} - \frac{1}{2\rho} \vec{\nabla}_{\zeta}^{2} - \frac{1}{2M} \vec{\nabla}_{R}^{2},
\end{eqnarray}
where we define the reduced masses 
\begin{eqnarray}
\mu &=& \frac{m_1 m_2}{m_{12}}, \\
\nu &=& \frac{m_{12}m_3}{m_{123}}, \\
\rho &=& \frac{m_{123}m_4}{M}.
\end{eqnarray}
The potential term is also rewritten as
\begin{eqnarray}
V 
 &=& \frac{k}{2}
\left\{
\left( \frac{1}{2} + \frac{1}{2}\frac{m_1^2+m_2^2}{m_{12}^2} \right) \vec{\xi}^{\,\,2}
+ \left( \frac{1}{2} + \frac{1}{2} \frac{m_{12}^2+m_3^2}{m_{123}^2} \right) \vec{\eta}^{\,\,2}
+ \vec{\zeta}^{\,\,2} \right. \nonumber \\
&& \left.
+ \frac{1}{2} \left( \frac{-m_1 + m_2}{m_{12}} + \frac{-m_1 + m_2}{m_{12}} \frac{m_3}{m_{123}} \right) \vec{\xi} \cdot \vec{\zeta} \right. \nonumber \\
&& \left.
+ \frac{1}{2} \frac{-m_1+m_2}{m_{12}} \vec{\xi} \cdot \vec{\zeta} 
+ \left( \frac{-m_1 - m_2 + m_3}{m_{123}} \right) \vec{\eta} \cdot \vec{\zeta}
\right\}.
\end{eqnarray}
Interestingly, for $m_1=m_2$, the relative coordinate $\vec{\xi}$ is decoupled from $\vec{\eta}$ and $\vec{\zeta}$ completely.\footnote{There is still a coupling for $\vec{\eta}$ and $\vec{\zeta}$. However, this coupling is irrelevant to the estimation of the wave function of ${\rm c}{\rm c}$ quark pair as discussed below.}
This is the special property for the harmonic oscillator potential in Eq.~(\ref{eq:HO_potential}).
For the $\mathrm{cc}$ pair (assigned as the particle 1 and 2) described by $\vec{\xi}$, we obtain the Hamiltonian for $\vec{\xi}$
\begin{eqnarray}
H_{\xi}^{\Tcc\three} = -\frac{1}{2\mu} \vec{\nabla}_{\xi}^2 + \frac{3}{4} \frac{k}{2} \, \vec{\xi}^{\,\,2},
\label{eq:Tcc3_hamiltonian}
\end{eqnarray}
which happens to be the same Hamiltonian for the $\mathrm{cc}$ pair in $\Xi_{\mathrm{cc}}$.\footnote{In case of $\Xi_{{\rm c}{\rm c}}$, we consider the three-body system ${\rm c}{\rm c}{\rm q}$ and find that the Hamiltonian for ${\rm c}{\rm c}$ quark pair is given by
\begin{eqnarray}
H_{\xi}^{\Xi_{{\rm c}{\rm c}}} = -\frac{1}{2\mu} \vec{\nabla}_{\xi}^{2} + \frac{3}{4} \frac{k}{2} \, \vec{\xi}^{\,\,2},
\end{eqnarray}
with the relative coordinate $\vec{\xi}$ for ${\rm c}{\rm c}$ distance, which turns out to be the same as Eq.~(\ref{eq:Tcc3_hamiltonian}).
}
Consequently, we obtain the value of the wave function at the center-of-mass of the $\mathrm{cc}$ pair in $\Tcc\three$
\begin{eqnarray}
R_{{\rm c}{\rm c}}^{\Tcc\three}(0) = \frac{2}{b^{3/2}} = 1.06 \hspace{0.5em}\mathrm{GeV}^{3/2},
\end{eqnarray}
with $b=1/\sqrt{\mu\,\omega}$ and $\omega = \sqrt{3/4}\sqrt{k/\mu}$ for $k=0.331$ GeV$^{3}$ and $\mu=m_{\mathrm c}/2=1.48/2$ GeV.

\subsubsection{$\Tcc\six$}
\label{sec:Tcc6}
Next, we consider the case that the pair of two quarks (antiquarks) belongs to ${\bf 6}_{\mathrm c}$.
Using the values of $\vec{\lambda}_{i} \cdot \vec{\lambda}_{j}$ ($i < j$) as
\begin{eqnarray}
\vec{\lambda}_{1} \cdot \vec{\lambda}_{2} &=& \frac{4}{3}, \nonumber \\
\vec{\lambda}_{1} \cdot \vec{\lambda}_{3} &=& \frac{2}{3} \left( -\frac{16}{3} \right) + \frac{1}{3} \frac{2}{3} = -\frac{10}{3}, \nonumber \\
\vec{\lambda}_{1} \cdot \vec{\lambda}_{4} &=& \frac{2}{3} \left( -\frac{16}{3} \right) + \frac{1}{3} \frac{2}{3} = -\frac{10}{3}, \nonumber \\
\vec{\lambda}_{2} \cdot \vec{\lambda}_{3} &=& \frac{2}{3} \left( -\frac{16}{3} \right) + \frac{1}{3} \frac{2}{3} = -\frac{10}{3}, \nonumber \\
\vec{\lambda}_{2} \cdot \vec{\lambda}_{4} &=& \frac{2}{3} \left( -\frac{16}{3} \right) + \frac{1}{3} \frac{2}{3} = -\frac{10}{3}, \nonumber \\
\vec{\lambda}_{3} \cdot \vec{\lambda}_{4} &=& \frac{4}{3},
\end{eqnarray}
we obtain the potential given by
\begin{eqnarray}
V 
&=&
\frac{k}{2}
\left\{
\left( -\frac{1}{4} + \frac{5}{4} \frac{m_1^2+m_2^2}{m_{12}^2} \right) \vec{\xi}^{\,\,2}
+ \left( \frac{5}{4} + \frac{1}{4} \frac{-m_{12}^2+5m_3^2}{m_{123}^2} \right) \vec{\eta}^{\,\,2}
+ \vec{\zeta}^{\,\,2}
\right. \nonumber \\
&& \left.
+ 2 \left( \frac{5}{8} \frac{-m_1+m_2}{m_{12}} + \frac{5}{8} \frac{-m_1+m_2}{m_{12}} \frac{m_3}{m_{123}} \right) \vec{\xi} \cdot \vec{\eta}
\right. \nonumber \\
&& \left.
+ 2 \frac{5}{8} \frac{-m_1+m_2}{m_{12}} \vec{\xi} \cdot \vec{\zeta}
+ 2 \frac{1}{4} \frac{m_{12}+5m_3}{m_{123}} \vec{\eta} \cdot \vec{\zeta}
\right\},
\end{eqnarray}
where the relative coordinate in Eq.~(\ref{eq:relative_4_a}) is used.
For the case of $m_1=m_2$,
the relative coordinate $\vec{\xi}$ for the $\mathrm{cc}$ pair is decoupled from $\vec{\eta}$ and $\vec{\zeta}$ decoupled again,
and hence we obtain the Hamiltonian
\begin{eqnarray}
H_{\xi}^{\Tcc\six} = -\frac{1}{2\mu} \vec{\nabla}_{\xi}^2 + \frac{3}{8} \frac{k}{2} \, \vec{\xi}^{\,\,2},
\label{eq:Tcc6_hamiltonian}
\end{eqnarray}
for the $\mathrm{cc}$ pair.
Consequently, we get the value of the wave function at the center-of-mass of the $\mathrm{cc}$ pair in ${\mathrm T}_{\mathrm{cc}}({\bf 6}_{\mathrm c})$
\begin{eqnarray}
R_{{\rm c}{\rm c}}^{\Tcc\six}(0) = \frac{2}{b^{3/2}} = 0.817 \hspace{0.5em}\mathrm{GeV}^{3/2},
\end{eqnarray}
with $b=1/\sqrt{\mu\,\omega}$ and $\omega = \sqrt{3/8}\sqrt{k/\mu}$ for $k=0.331$ GeV$^{3}$ and $\mu=m_{\mathrm c}/2=1.48/2$ GeV (Set A).
The value for $\Tcc\six$ is smaller than that for $\Tcc\three$.
This is indeed the case, because the potential strength in Eq.~(\ref{eq:Tcc6_hamiltonian}) is half as small as that in Eq.~(\ref{eq:Tcc3_hamiltonian}).
We show also the result for $k=0.0577$ GeV$^{3}$ and $\mu=m_{\rm c}/2=1.326$ GeV from the mass splittings of $s$-wave and $p$-wave charmonia (Set B).
The summary of the present result for $R_{{\rm cc}}(0)$ is given in Table.~\ref{table:R0}.
Those values will be used in the estimation of the $\Tcc$ production in the next section.

\begin{table}[tbp]
\caption{The values of the wave function at the center-of-mass $R_{{\rm cc}}(0)$ for the $\mathrm{cc}$ pair in $\Tcc\three$ and $\Tcc\six$, respectively. Units are $\mathrm{GeV}^{3/2}$. }
\begin{center}
\renewcommand{\arraystretch}{1.2}
\begin{tabular}{c|c|c}
\hline\hline
 & Set A & Set B  \\
 \hline
 $\Tcc\three$ & 1.06  & 0.53 \\
\hline
 $\Tcc\six$ & 0.82 & 0.41 \\
\hline\hline
\end{tabular}
\end{center}
\label{table:R0}
\end{table}%

\section{Inclusive production of tetraquark $\Tcc$}
\label{sec:production}

To pin down the properties of exotic color configurations of $\Tcc$, it is important to relate the internal structure of $\Tcc$ to some experimental observables. Because the $\Tcc$ production requires at least two ${\rm c}\bar{\rm c}$ pairs, low energy exclusive production is not very promising. Refs.~\cite{Cho:2010db,Cho:2011ew,Cho:2017dcy} have discussed the production of $\Tcc$ in heavy ion collisions where charm quark pairs are produced abundantly. Here we consider the inclusive $\Tcc$ production in electron-positron collisions. In fact, the double-charm productions with charmonia in the final state are analyzed at Belle and BaBar~\cite{Abe:2002rb,Abe:2004ww,Aubert:2005tj}. In this section, we introduce the theoretical framework to describe the $\Tcc$ production.

\subsection{NRQCD framework}

To evaluate the production of $\Tcc$, we utilize the non-relativistic QCD (NRQCD)~\cite{Bodwin:1994jh,Petrelli:1997ge}, which is an effective field theory for heavy quarks. This framework allows one to factorize the hard part of the process in which the heavy quarks are produced at short distance and the soft part where the heavy quarks fragment into hadrons. The hard process is calculable through the matching of the NRQCD operator with the perturbative QCD amplitude as an expansion of $\alpha_{s}$. The soft part is given by the nonperturbative matrix elements of the NRQCD operators expanded in powers of the heavy quark velocity $v$. 

NRQCD was originally introduced to describe the annihilation decay and the production of single charmonium, and was later applied to double-charm productions with charmonia in the final states~\cite{Braaten:2002fi,Liu:2002wq,Zhang:2005cha,Bodwin:2007ga,Zhang:2008gp}. 
It was shown that the higher order corrections both in $\alpha_{s}$~\cite{Zhang:2008gp} and velocity expansion~\cite{Bodwin:2007ga} are important for the double-charmonium productions. 

For the inclusive $\Tcc$ production in electron-positron collisions (${\rm e}^{+}{\rm e}^{-}\to \Tcc[\alpha]+{\rm X}$), the differential cross section is given by
\begin{align}
    &{\rm d}\sigma_{\alpha}({\rm e}^{+}{\rm e}^{-}\to \Tcc[\alpha]+{\rm X}) 
    \nonumber \\
    =&
    \sum_{k}
    {\rm d}\hat{\sigma}({\rm e}^{+}{\rm e}^{-}\to [{\rm cc}]^{k}_{\alpha}+\bar{\rm c}+\bar{\rm c})
     \langle \mathcal{O}^{k}(\Tcc [\alpha])\rangle \nonumber \\
    =&
    \sum_{k}
    {\rm d}\hat{\sigma}({\rm e}^{+}{\rm e}^{-}\to [{\rm cc}]^{k}_{\alpha}+\bar{\rm c}+\bar{\rm c})
    |\bra{ \Tcc[\alpha]+{\rm X}^{\prime}} [{\rm cc}]^{k}_{\alpha}
    \ket{0}|^{2},
\end{align}
where $\alpha$ stands for the color-spin configuration of cc pair inside T$_{\rm cc}$, $k$ indicates the order of the velocity expansion, and ${\rm X}^{\prime}$ represents the light components nonperturbatively produced during the soft hadronization process of the cc pair to the $\Tcc$ state. 
${\rm d}\hat{\sigma}$ is the cross section for the elementary process ${\rm e}^{+}{\rm e}^{-}\to [{\rm cc}]^{k}_{\alpha}+\bar{\rm c}+\bar{\rm c}$, where the ${\rm c}{\rm c}$ pair is combined to form the color-spin configuration $\alpha$ at order $k$.
As a first trial, here we consider the leading order calculation both in $\alpha_{s}$ and $v$. In this case, the nonperturbative matrix element is a number for each $\alpha$ and we denote it as
\begin{align}
    |\bra{\Tcc[\alpha]+{\rm X}^{\prime}} [{\rm cc}]^{k}_{\alpha}
    \ket{0}|^{2}
    \Bigr|_{k={\rm LO}}
    =&
    \begin{cases}
    h_{[\bm{\bar{3}},{}^{3}{\rm S}_{1}]} & \text{for } 
    \alpha = [\bm{\bar{3}},{}^{3}{\rm S}_{1}] ,\\
    h_{[\bm{6},{}^{1}{\rm S}_{0}]} & \text{for } 
    \alpha = [\bm{6},{}^{1}{\rm S}_{0}] .
    \end{cases}
\label{eq:np_matrix_elements}
\end{align}
In the following, we first present the kinematics of the reaction in Section~\ref{subsec:kinematics}, and then show the calculation of the hard process in Section~\ref{subsec:hard}. The treatment of the soft matrix element is discussed in Section~\ref{subsec:soft}.

\subsection{Kinematics}\label{subsec:kinematics}

\begin{figure}[tbp]
\centering
\vspace*{-8em}
\hspace*{-2em}
\includegraphics[width=16cm,clip]{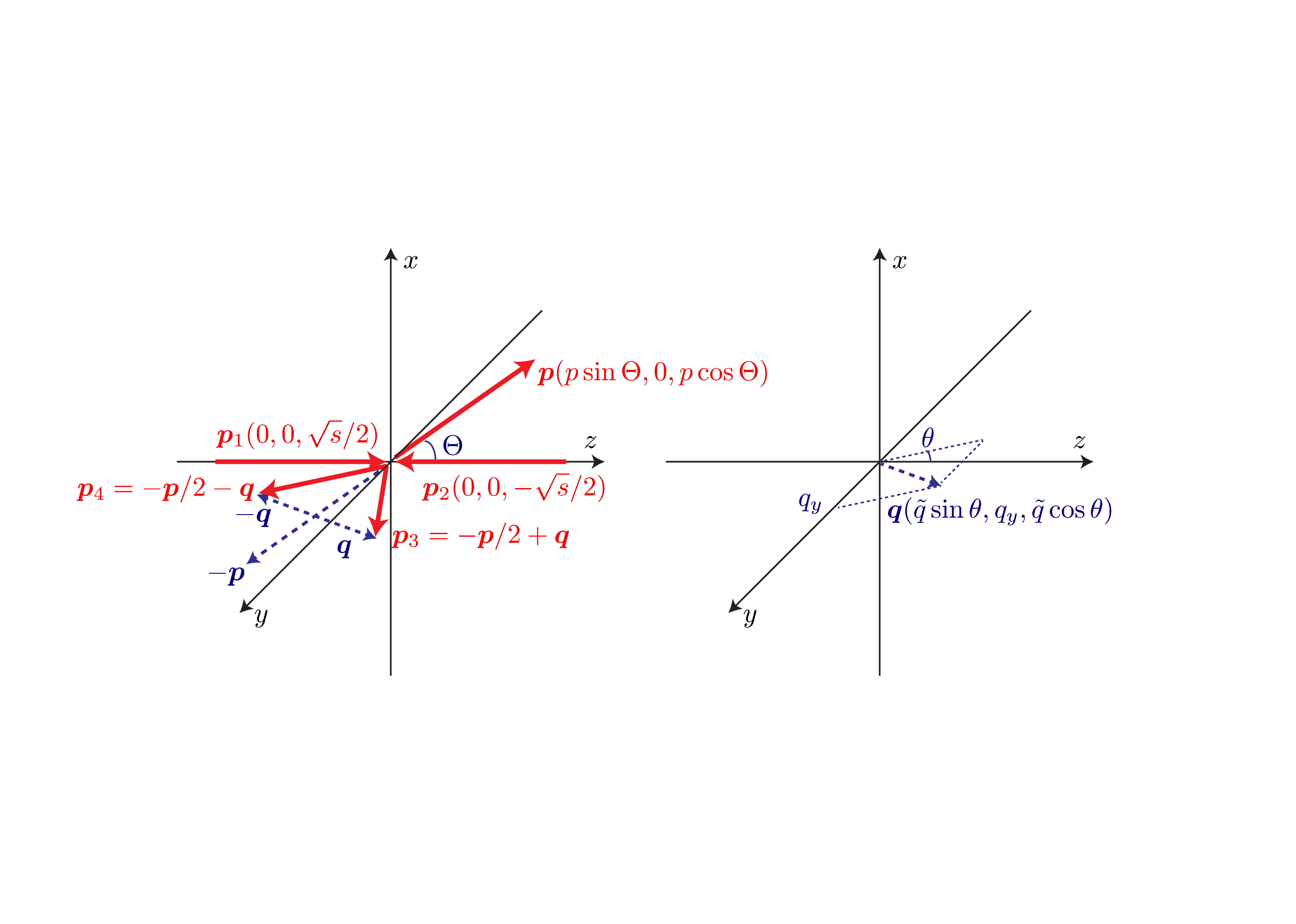}
\vspace*{-8em}
\caption{\label{fig:kinematics} Assignment of the momentum variables of the inclusive $\Tcc$ production in the center of mass frame.}
\end{figure}

We assign momentum variables as ${\rm e}^{+}(p_{1}){\rm e}^{-}(p_{2})\to [{\rm cc}]^{k}_{\alpha}(p)+\bar{\rm c}(p_{3})+\bar{\rm c}(p_{4})$. We choose the collision axis as the $z$ direction in the center of mass frame, and $x$ axis is set so that $\Tcc$ moves in the $xz$ plane. We write the magnitude of the three-momentum $\bm{p}$ as $p$ and the angle of $\bm{p}$ from the $z$ axis as $\Theta$. The momenta of anticharm quarks are given by $\bm{p}_{3}=-\bm{p}/2+\bm{q}$ and $\bm{p}_{4}=-\bm{p}/2-\bm{q}$ with $\bm{q}=(\tilde{q}\sin\theta,q_{y},\tilde{q}\cos\theta)$ in the cylindrical polar coordinates. These momenta and the coordinate system are summarized in Fig.~\ref{fig:kinematics}. The four velocities of the system are then written as
\begin{align}
    p_{1}^{\mu}
    =& (E_{1},0,0,\tfrac{\sqrt{s}}{2}), \\
    p_{2}^{\mu}
    =& (E_{2},0,0,-\tfrac{\sqrt{s}}{2}), \\
    p^{\mu}
    =&
    (E_{p},p\sin\Theta,0,p\cos\Theta) , \\
    p_{3}^{\mu}
    =& (E_{3},-\tfrac{p}{2}\sin\Theta+\tilde{q}\sin\theta,
    q_{y},-\tfrac{p}{2}\cos\Theta+\tilde{q}\cos\theta), \\
    p_{4}^{\mu}
    =& (E_{4},-\tfrac{p}{2}\sin\Theta-\tilde{q}\sin\theta,
    -q_{y},-\tfrac{p}{2}\cos\Theta-\tilde{q}\cos\theta),
\end{align}
where $s$ is the total energy squared. We neglect the electron mass, and regard the mass of $\Tcc$ as $2m_{\rm c}$ in the leading order of the velocity expansion. The energies are then given by
\begin{align}
    E_{1}
    =&\tfrac{\sqrt{s}}{2}, \\
    E_{2}
    =&\tfrac{\sqrt{s}}{2}, \\
    E_{p}
    =&\sqrt{4m_{\rm c}^{2}+p^{2}}, \\
    E_{3}
    =&\sqrt{m_{\rm c}^{2}+\tfrac{p^{2}}{4}+\tilde{q}^{2}+q_{y}^{2}
    -p\tilde{q}\cos(\theta-\Theta)}, \\
    E_{4}
    =&\sqrt{m_{\rm c}^{2}+\tfrac{p^{2}}{4}+\tilde{q}^{2}+q_{y}^{2}
    +p\tilde{q}\cos(\theta-\Theta)},
\end{align}

With the above setup, the total cross section is given by
\begin{align}
    \sigma_{\alpha}
    =&
    \frac{1}{2}
    \frac{1}{2s}\frac{1}{m_{\rm c}}\int \frac{d^{3}p}{(2\pi)^{3}}\frac{1}{2E_{p}}
    \frac{d^{3}p_{3}}{(2\pi)^{3}}\frac{1}{2E_{3}}
    \frac{d^{3}p_{4}}{(2\pi)^{3}}\frac{1}{2E_{4}} \nonumber \\
    &\times (2\pi)^{4}\delta^{4}(p_{1}+p_{2}-p-p_{3}-p_{4})\overline{\sum}|\hat{\mathcal{M}}_{\alpha}|^{2} h_{\alpha},
\end{align}
where $\overline{\sum}|\hat{\mathcal{M}}_{\alpha}|^{2}$ stands for the color-spin projected squared amplitude with the summation over the final spins and average over the initial spins.
The factor 1/2 is the symmetry factor of identical $\bar{c}$ quarks in the final state, and $1/m_{\rm c}$ comes from the non-relativistic normalization of the state vector in the matrix element $h_{\alpha}$. Performing the phase space integration, we obtain the differential cross section with respect to the $\Tcc$ momentum $p$ and scattered angle $\cos\Theta$ as
\begin{align}
    \frac{{\rm d}\sigma_{\alpha}}{{\rm d}p\ {\rm d}\cos\Theta}
    =&
    \frac{1}{(2\pi)^{4}}\frac{p^{2}}{16sE_{p}m_{\rm c}}
    \int_{0}^{2\pi}{\rm d}\theta
    \int_{0}^{\tilde{q}_{\text{max}}}{\rm d}\tilde{q} 
    \frac{\tilde{q}}{q_{y}(E_{3}+E_{4})}
    \overline{\sum}|\hat{\mathcal{M}}_{\alpha}|^{2}h_{\alpha}
    \label{eq:difcross} ,
\end{align} 
where we define
\begin{align}
    q_{y}
    =&
    \frac{
    \sqrt{
    A
    -B\tilde{q}^{2}
    +C\tilde{q}^{2}\cos^{2}(\theta-\Theta)}}{2(\sqrt{s}-E_{p})}
    , \\
    \tilde{q}_{\text{max}}
    =&\sqrt{\frac{A}{B-C \cos^{2}(\theta-\Theta)}}
    , 
\end{align}
with $A = \sqrt{s}(\sqrt{s}-2E_{p})(\sqrt{s}-E_{p})^{2}$, $B=4(\sqrt{s}-E_{p})^{2}$, and $C=4p^{2}$. Although it seems that the integrand diverges ($q_{y}\to 0$) when $\tilde{q}\to \tilde{q}_{\rm max}$, this is not a real singularity and can be removed by the transformation of the integration variable as $x^{2}=\tilde{q}_{\rm max}-\tilde{q}$ in practice.

\subsection{Perturbative QCD process}\label{subsec:hard}

\begin{figure}[tbp]
\centering
\vspace*{-14em}
\hspace*{-9em}
\includegraphics[width=22cm,clip]{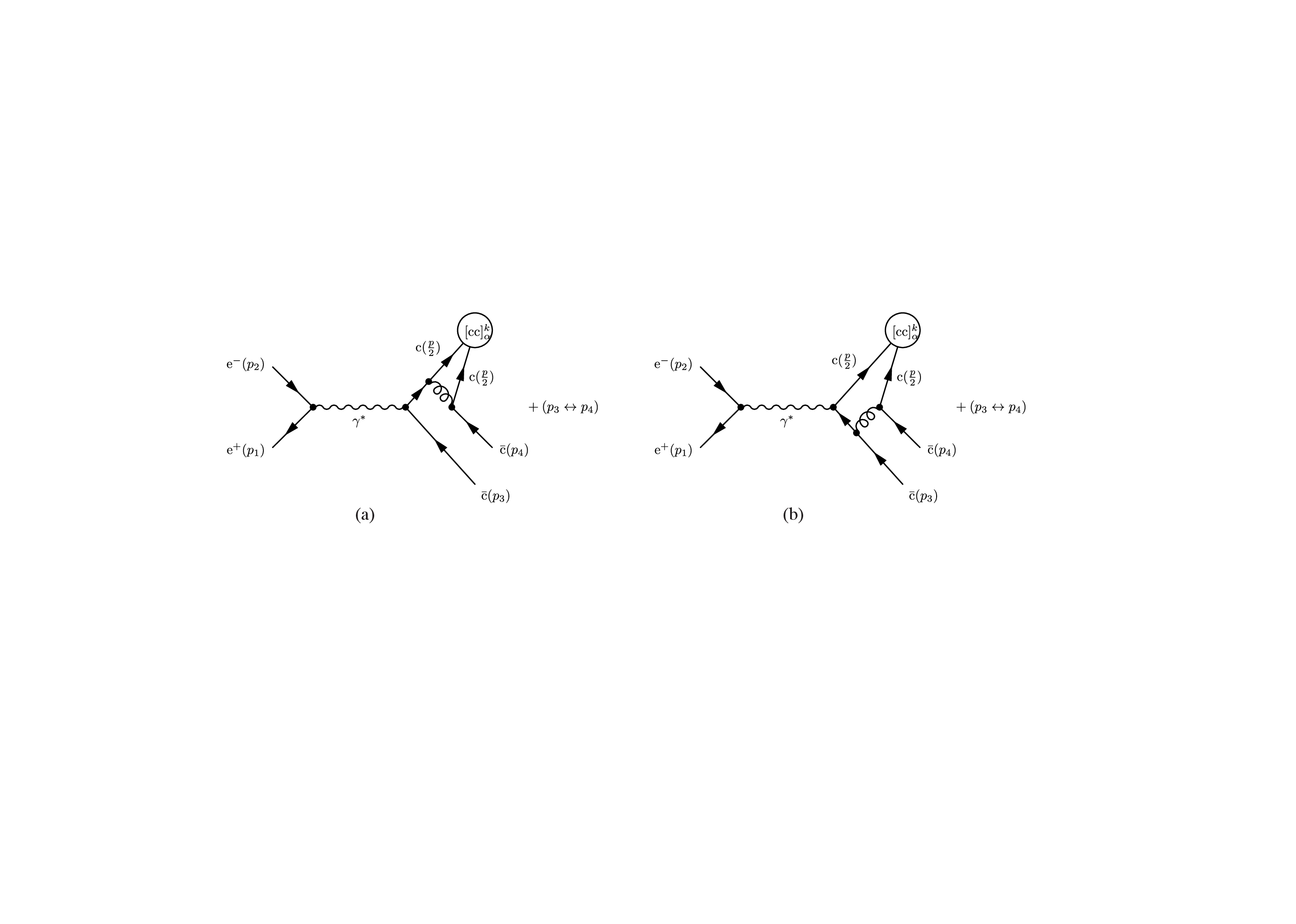}
\vspace*{-18em}
\caption{\label{fig:Diagrams} Feynman diagrams for the leading order contributions to the perturbative ${\rm d}\hat{\sigma}({\rm e}^{+}{\rm e}^{-}\to [{\rm cc}]^{k}_{\alpha}+\bar{\rm c}+\bar{\rm c})$ process.}
\end{figure}

In the leading order of $\alpha_{s}$, the perturbative amplitude for the ${\rm e}^{+}{\rm e}^{-}\to [\rm{cc}]^{k}_{\alpha}+\bar{\rm c}+\bar{\rm c}$ process $\hat{\mathcal{M}}_{\alpha}$ is expressed by four diagrams shown in Fig.~\ref{fig:Diagrams}. The explicit forms of these amplitudes are given by
\begin{align}
    -i\hat{\mathcal{M}}_{\alpha}
    =&
    -i\hat{\mathcal{M}}^{(a)}_{\alpha}
    -i\hat{\mathcal{M}}^{(a^{\prime})}_{\alpha}
    -i\hat{\mathcal{M}}^{(b)}_{\alpha}
    -i\hat{\mathcal{M}}^{(b^{\prime})}_{\alpha},
\end{align}
with
\begin{align}
    -i\hat{\mathcal{M}}_{\alpha}^{(a)}
    =&
    \bar{v}_{\rm e}(p_{1})(ie\gamma^{\mu})u_{\rm e}(p_{2})
    \frac{-ig_{\mu\nu}}{(p_{1}+p_{2})^{2}}
    \nonumber \\
    &\times
    v_{l}^{t}(p_{3})
    (-ie_{\rm c}(\gamma^{\nu})^{t})
    \left(\frac{i}{\Slash{p}+\Slash{p}_{4}-m_{\rm c}}\right)^{t}
    (-ig(\gamma^{\rho})^{t}(T^{a})^{t}_{lk}) \nonumber \\
    &\times [P_{\alpha,m}]_{kj}
    \frac{-ig_{\rho\sigma}\delta^{ab}}{(p/2+p_{4})^{2}}
    (-ig\gamma^{\sigma})
    (T^{b})_{ji}
    v_{i}(p_{4}),
     \\
    -i\hat{\mathcal{M}}^{(a^{\prime})}_{\alpha}
    =&
    \bar{v}_{\rm e}(p_{1})(ie\gamma^{\mu})u_{\rm e}(p_{2})
    \frac{-ig_{\mu\nu}}{(p_{1}+p_{2})^{2}} 
    \nonumber \\
    &\times
    v_{i}^{t}(p_{4})
    (-ie_{\rm c}\gamma^{\nu})^{t}
    \left(\frac{i}{\Slash{p}+\Slash{p}_{3}-m_{\rm c}}\right)^{t}
    (-ig\gamma^{\rho})^{t}(T^{a})^{t}_{ik} \nonumber \\
    &\times [P_{\alpha,m}]_{kj}
    \frac{-i\delta^{ab}g_{\rho\sigma}}{(p/2+p_{3})^{2}}
    (ig\gamma^{\sigma})
    (T^{b})_{jl}
    v_{l}(p_{3}),
     \\
    -i\hat{\mathcal{M}}^{(b)}_{\alpha}
    =&
    iee_{\rm c}g^{2}\bar{v}_{\rm e}(p_{1})\gamma^{\mu}u_{\rm e}(p_{2})
    \frac{g_{\mu\nu}}{(p_{1}+p_{2})^{2}} 
    \nonumber \\
    &\times
    v_{l}^{t}(p_{3})
    (\gamma^{\rho})^{t}
    (T^{a})^{t}_{lk}
    \left(\frac{1}{-\Slash{p}/2-\Slash{p}_{3}-\Slash{p}_{4}-m_{\rm c}}\right)^{t}
    (\gamma^{\nu})^{t}
    \nonumber \\
    &\times
    [P_{\alpha,m}]_{kj}
    \frac{\delta^{ab}g_{\rho\sigma}}{(p/2+p_{4})^{2}}
    \gamma^{\sigma}
    (T^{b})_{ji}
    v_{i}(p_{4}),
    \\
    -i\hat{\mathcal{M}}^{(b^{\prime})}_{\alpha}
    =&
    -iee_{\rm c}g^{2}\bar{v}_{\rm e}(p_{1})\gamma^{\mu}u_{\rm e}(p_{2})
    \frac{g_{\mu\nu}}{(p_{1}+p_{2})^{2}} 
    \nonumber\\
    &\times
    v_{i}^{t}(p_{4})
    (\gamma^{\rho})^{t}
    (T^{a})^{t}_{lk}
    \left(\frac{1}{-\Slash{p}/2-\Slash{p}_{3}-\Slash{p}_{4}-m_{\rm c}}\right)^{t}
    (\gamma^{\nu})^{t}
    \nonumber \\
    &\times
    [P_{\alpha,m}]_{kj}
    \frac{\delta^{ab}g_{\rho\sigma}}{(p/2+p_{3})^{2}}
    \gamma^{\sigma}
    (T^{b})_{jl}
    v_{l}(p_{3}),
\end{align}
where $u_{\rm e}$, $v_{\rm e}$ are the electron and positron fields, $u_{i}$, $v_{i}$ are the charm quark and antiquark fields with color $i$, and $e_{\rm c}=2e/3$ is the charge of the charm quark. The color-spin projection operators $P_{\alpha,m}$ are defined as 
\begin{align}
    P^{(\lambda)}_{\bm{\bar{3}},m}
    =&\sum_{\bm{\bar{3}},{}^{3}{\rm S}_{1}}
    \bar{u}_{k}^{t}(p/2)
    \bar{u}_{j}(p/2)
    =\frac{1}{\sqrt{2}}
    \left(\frac{\Slash{p}}{2}+m_{\rm c}\right)^{t}
    \Slash{\epsilon}^{(\lambda)t}C\Phi_{mkj}^{\rm A}
    , \label{eq:projection3bar} \\
    P_{\bm{6},m}
    =&\sum_{\bm{6},{}^{1}{\rm S}_{0}}
    \bar{u}_{k}^{t}(p/2)
    \bar{u}_{j}(p/2)
    =\frac{1}{\sqrt{2}}
    \left(\frac{\Slash{p}}{2}+m_{\rm c}\right)^{t}
    \gamma_{5}C\Phi_{mkj}^{\rm S}
    ,  \label{eq:projection6}
\end{align}
where $C$ is the charge conjugation matrix, $\epsilon^{(\lambda)}_{\mu}$ is the polarization vector of spin triplet $\Tcc$, and $\Phi_{mkj}^{\rm A,S}=\mp\Phi_{mjk}^{\rm A,S}$ are the normalized tensors in color space.\footnote{In Ref.~\cite{Hyodo:2012pm}, transpose of the $(\Slash{p}/2+m_{\rm c})$ factor has been missed as typos.} The suffix $m$ specifies the color of $\Tcc$, which runs 1-3 (4-9) for color $\bm{\bar{3}}$ ($\bm{6}$). The antisymmetric part is related to the Levi-Civita symbol $\Phi_{mkj}^{\rm A}=\epsilon_{mkj}/\sqrt{2}$. The factor $1/\sqrt{m_{\rm c}}$ is introduced to account for the normalization of the state vector $\ket{\Tcc[\alpha]+{\rm X}_{N}}$. 

Calculating the squared amplitude, we obtain the expression with $m_{\rm c}$ and products of $p_{1}$, $p_{2}$, $p$, $p_{3}$ and $p_{4}$. For instance, we have 
\begin{eqnarray}
   \overline{\sum}|\hat{\mathcal{M}}_{\alpha}^{(a)}|^{2}
   &=&\cfrac{2 e^2 e_{\rm c}^2 g^4 }
   {3m_{\rm c}
   (p_{1}\cdot p_{2})^{2} \left(2m_{\rm c}^2+p\cdot p_{4}\right)^4} 
   \Bigl[4 m_{\rm c}^4 p\cdot p_{1}\
          p_{2}\cdot p_{3} \nonumber \\
   &&
          +4 m_{\rm c}^2 p\cdot p_{2}\
          p_{1}\cdot p_{3} 
          (m_{\rm c}^2+p\cdot p_{4}) 
          +4m_{\rm c}^2 p\cdot p_{1}\
          p\cdot p_{4} \
          p_{2}\cdot p_{3} \nonumber \\
   &&
          +4 m_{\rm c}^2 p\cdot p_{4}\
          p_{1}\cdot p_{4}\
          p_{2}\cdot p_{3}
          +4 m_{\rm c}^2 p\cdot p_{4}\
          p_{1}\cdot p_{3}\
          p_{2}\cdot p_{4} \nonumber \\
   &&      -p_{1}\cdot p_{2} 
          \left(8 m_{\rm c}^6+8 m_{\rm c}^4 p\cdot p_{4}
          +3m_{\rm c}^2 (p\cdot p_{4})^{2}\right) \nonumber \\
   &&      
          +(p\cdot p_{4})^{2}
          p_{1}\cdot p_{4} \
          p_{2}\cdot p_{3}
          +(p\cdot p_{4})^{2}
          p_{1}\cdot p_{3}\
          p_{2}\cdot p_{4}\Bigr],
\end{eqnarray}
for $[\alpha]=[\bm{6},{}^{1}{\rm S}_{0}]$ case. Using the kinematics given in Section~\ref{subsec:kinematics}, we can write the squared amplitude in terms of $\sqrt{s},p,\Theta,\tilde{q},q_{y},\theta$, which is substituted into Eq.~\eqref{eq:difcross} to calculate the differential cross sections.

\subsection{Nonperturbative matrix element}\label{subsec:soft}

In order to estimate the absolute values of the production cross sections in Eq.~(\ref{eq:np_matrix_elements}),
we need to evaluate the soft matrix elements, which describe ``fragmentation'' of cc into $\Tcc$.

So far, we have computed the hard part of the reaction, which produces charm quarks, {\it i.e.},
${\rm e}^+ + {\rm e}^- \to  {\rm cc} + \hbox{(anything)}$.
The hard part is followed by soft hadronization processes. 
We here assume that the soft parts can be factorized and are given by matrix elements of final-state
hadrons. 
For heavy quarks, one may regard the soft part as
the probability of finding two charm quarks at the origin in the hadron,
i.e., $\propto |\psi_{\rm cc}(0)|^2$, where $\psi_{{\rm cc}}(0)$ is the relative wave function of cc in $\Tcc$.
Thus, we may use the expectation value at center-of-mass for the wave functions of the ${\rm c}{\rm c}$ quark pairs in $\Tcc$ by assigning $h_{[\bm{\bar{3}},{}^{3}{\rm S}_{1}]}$ and $h_{[\bm{6},{}^{1}{\rm S}_{0}]}$ to $| R_{{\rm c}{\rm c}}^{\Tcc\three}(0) |^2/4\pi$ and $| R_{{\rm c}{\rm c}}^{\Tcc\six}(0) |^2/4\pi$, respectively,
as introduced in our previous work \cite{Hyodo:2012pm}.
However, this is not enough for qualitative estimations, because no dynamical information to pick up light quarks from vacuum is involved.
Otherwise, if one supposed no extra factor for light quarks in forming $\Tcc$, then
the production rates of $\Xi_{\rm cc}$ and $\Tcc$ would be of the same order.
Namely, it is as easy to produce $\Tcc$ as $\Xi_{\rm cc}$, although $\Tcc$ requires more light quarks, (u and d), than $\Xi_{\rm cc}$ does.
This observation clearly contradicts with our expectations. 
Including the probability of picking up light quarks to form a hadron $X$, $\hbox{Prob (cc}\to X)$, we now estimate the matrix elements as
\begin{align}
    h_{[\bm{\bar{3}},{}^{3}{\rm S}_{1}]}
    &=\hbox{Prob (cc}\to \Xi_{\rm cc} )
    \times
    | R_{{\rm c}{\rm c}}^{\Xi_{\rm cc}}(0) |^2/4\pi , \\
    h_{[\bm{6},{}^{1}{\rm S}_{0}]}
    &=\hbox{Prob (cc}\to \Tcc\three) 
    \times
    | R_{{\rm c}{\rm c}}^{\Tcc\three}(0) |^2/4\pi , \\
    h_{[\bm{6},{}^{1}{\rm S}_{0}]}
    &=\hbox{Prob (cc}\to \Tcc\six) 
    \times
    | R_{{\rm c}{\rm c}}^{\Tcc\six}(0) |^2/4\pi .
\end{align}

We thus need to take into account the probability of picking up light quarks (from vacuum) to form the final hadron.
A similar situation is found in the production of heavy baryons and mesons, and it is useful for us to investigate the situations for those known hadrons.
Let us compare the productions of D/B $({\rm Q}\bar{\rm q})$ with those of $\Lambda_{\rm Q}$, $\Sigma_{\rm Q}$ (Qqq), because the
analogous investigation can be applied for $\Xi_{\rm QQ}$ (QQq) and ${\rm T}_{\rm QQ}$ (${\rm Q}^2\bar{\rm q}^2$) by regarding ${\rm QQ}$ as ${\rm Q}$.
High energy production of b quark is followed by the fragmentation into B mesons and bottom baryons, 
eventually decaying via the weak interaction.
Fragmentation probabilities of ${\rm B}_{\rm u}$, ${\rm B}_{\rm d}$, ${\rm B}_{\rm s}$, $\Lambda_{\rm b}$ productions were measured 
at LEP (Z decay) \cite{Abreu:1995me,Barate:1997if}, CDF (p$\bar{\rm p}$) \cite{Affolder:1999iq,Aaltonen:2008zd}, and LHCb (pp) \cite{Aaij:2011jp}.

The most recent results from the LHCb~\cite{Aaij:2011jp} show the ratios,
\begin{align}
\frac{f_{\rm ud}}{f_{\rm u}+f_{\rm d}} 
&= (0.404 \pm 0.017({\rm stat})\pm 0.027({\rm syst})\pm 0.105({\rm Br}) \nonumber \\ 
&\quad \times (1 - (0.031\pm 0.004({\rm stat})\pm 0.003({\rm syst})\times p_{\rm T} \hbox{[GeV]}),
\end{align}
where $f_{\rm u, d, ud}$ are probabilities of b-quark fragmentation into ${\rm B}^-({\rm b}\bar{\rm u})$, ${\rm B}^0 ({\rm b}\bar{\rm d})$,
and $\Lambda_{\rm b}$(bud), respectively.
It should be noted that because the preceding strong decays are included, the above data includes the production probabilities from the excited states, ${\rm B}^*$ and $\Sigma_{\rm b}^{(*)}$ ($\Sigma_{\rm b}$ and $\Sigma_{\rm b}^{*}$), as well as the ground states, B and  $\Lambda_{\rm b}$:
\begin{eqnarray}
&& f_{\rm u} + f_{\rm d} = \hbox{Prob (} {\rm b}\to  {\rm B} + {\rm B}^* + \hbox{other excited states)}, \\
&& f_{\rm ud} = \hbox{Prob (} {\rm b}\to  \Lambda_{\rm b} + \Sigma_{\rm b} + \Sigma_{\rm b}^* + \hbox{other excited states)}, 
\end{eqnarray}
Measurements of the ${\rm b}$-quark fragmentation fractions in p$\bar{\rm p}$ collisions at $\sqrt{s} = 1.8$ TeV and 1.96 TeV
were reported by the CDF Collaboration~\cite{Aaltonen:2008zd},
giving
\begin{eqnarray}
&& \frac{f_{\rm ud}}{ f_{\rm u}+f_{\rm d}} = 
0.281 \pm 0.012({\rm stat})^{+0.058}_{-0.056}({\rm sys})^{+0.128}_{-0.087}(\mathcal{B}).
\end{eqnarray}

For understanding of ${\rm b}$-quark fragmentation,
here we heuristically assume the following simple model to calculate the ratios of production rates.
\begin{itemize}
\item[(1)] The leading heavy quark picks up light quarks and forms a heavy hadron. The light quarks are created as a quark-antiquark pair, whose ``probability''  is universal to all the processes.
\item[(2)] The hadrons are produced only when the heavy quarks meet light quarks
(or antiquarks) with the matched quantum numbers.
\end{itemize}
More intuitively,
q$\bar{\rm q}$ can be created in the color singlet, spin singlet $0^+$ state.
The pairs u$\bar{\rm u}$ and d$\bar{\rm d}$ are created with the same probability, while s$\bar{\rm s}$ may be partially suppressed due to the small breaking of flavor symmetry.

Now, let us suppose the probability of creating a u$\bar{\rm u}$ pair from the vacuum is $\eta_{\rm u}$.
Then the probability of creating $\bar{\rm u} (\bar R , \downarrow)$ with color anti-red and spin down is $(1/6)\eta_{\rm u}$ .
For a heavy quark Q$(R, \uparrow)$ to form a Q$\bar{\rm u}$ ($J^{\pi}=0^- $) meson, the probability is given by
$(1/3)(1/2)(1/6) \eta_{\rm u} = (1/36) \eta_{\rm u}$, where $1/3$ is the factor for the color singlet combination,
$1/2$ is for spin 0.
Similarly, for the vector Q$\bar{\rm u}$ ($1^-$) meson, it is $(1/3)(3/2)(1/6) \eta_{\rm u} =(1/12) \eta_u$.

We obtain the same probabilities for Q$\bar{\rm d}$ mesons, assuming $\eta_{\rm q} = \eta_{\rm u} = \eta_{\rm d}$, and, in total, the probability of producing
B $+ {\rm B}^*$ will be  
\begin{eqnarray}
&& f_{\rm u}+f_{\rm d} \sim  \frac{2}{9} \eta_{\rm q} .
\end{eqnarray}
Here we neglect the higher excited states in the total production rate.

Next, the production of baryons are estimated similarly by
\begin{eqnarray}
&& \hbox{Prob (}{\rm b}\to \Lambda_{\rm b} ) \sim \frac{1}{108} \eta_q^2,\\
&& \hbox{Prob (}{\rm b}\to\Sigma_{\rm b} )\sim \frac{1}{36} \eta_{\rm q}^2,\\
&& \hbox{Prob (}{\rm b}\to\Sigma_{\rm b}^* ) \sim  \frac{1}{18} \eta_{\rm q}^2,
\end{eqnarray}
and, neglecting the higher excited state, we obtain $f_{\rm B}$ as the sum,
\begin{eqnarray}
&& 
f_{\rm ud}= \frac{5}{54} \eta_{\rm q}^2.
\end{eqnarray}
Note that the baryon will be formed by picking two light quarks and then the probability is proportional to $\eta_{\rm q}^2$.

Thus our estimate of the production ratios of the heavy baryons and heavy mesons is 
\begin{eqnarray}
&& \frac{f_{\rm ud}}{f_{\rm u}+f_{\rm d}} = \frac{5}{12} \eta_{\rm q}. 
\end{eqnarray}
By using the experimental data for the ratio, $0.3\sim 0.4$, we estimate
\begin{eqnarray}
&& \eta_{\rm q} = 0.7\sim 1.0 .
\end{eqnarray}
In the numerical calculation, we adopt $\eta_{\rm q}=1$.

Once $\eta_{\rm q}$ is obtained, we may evaluate the production probabilities of $\Tcc$ and $\Xi_{\rm cc}$ in
a similar way.
We obtain the production probabilities of $\Xi_{\rm cc}$ ($J=1/2$) and  $\Xi^*_{\rm cc}$ ($J=3/2$) as 
\begin{eqnarray}
&& \hbox{Prob (cc}\to \Xi_{\rm cc} ) \sim \frac{1}{3}\,\frac{2}{3}\, \frac{1}{6}\, \eta_{\rm q}= \frac{1}{27} \,\eta_{\rm q},\\
&& \hbox{Prob (cc}\to \Xi^*_{\rm cc})\sim \frac{1}{3}\,\frac{4}{3}\,\frac{1}{6}\,\eta_{\rm q} = \frac{2}{27}\, \eta_{\rm q}.
\end{eqnarray} 
The production probabilities of $\Tcc$ are also given by
\begin{eqnarray}
&& \hbox{Prob (cc}\to \Tcc\three) \sim \frac{1}{108}\, \eta_{\rm q}^2,
\label{eq:Prob3}\\
&& \hbox{Prob (cc}\to \Tcc\six)\sim  \frac{1}{72}\, \eta_{\rm q}^2,
\label{eq:Prob6}
\end{eqnarray} 
for $I=0$ states and 
\begin{eqnarray}
&& \hbox{Prob (cc}\to \Tcc\three^{I=1}_{J}) \sim 
\begin{cases}
\frac{1}{324}\, \eta_{\rm q}^2, & J=0 \\
\frac{1}{108}\, \eta_{\rm q}^2, & J=1 \\
\frac{5}{324}\, \eta_{\rm q}^2, & J=2
\end{cases}
\label{eq:Prob3I1}\\
&& \hbox{Prob (cc}\to \Tcc\six^{I=1})\sim  \frac{1}{72}\, \eta_{\rm q}^2,
\label{eq:Prob6I1}
\end{eqnarray} 
for $I=1$ states.
Therefore, the ratio, $(\Tcc)/(\Xi_{\rm cc})$, is given by
\begin{eqnarray}
&& \frac{\hbox{Prob (cc}\to \Tcc\three)}{\hbox{Prob (cc}\to \Xi_{\rm cc})} 
 \sim \frac{1}{4} \,\eta_{\rm q} = 0.15 \sim 0.25,\\ 
&& \frac{\hbox{Prob (cc}\to \Tcc\six)}{\hbox{Prob (cc}\to \Xi_{\rm cc})}
 \sim \frac{3}{8} \,\eta_{\rm q} = 0.2 \sim 0.38.
\end{eqnarray}
Interestingly, we note that the ratio for productions of $\Tcc\three$ and $\Tcc\six$ is independent of $\eta_{\rm q}$;
\begin{eqnarray}
&& \frac{\hbox{Prob (cc}\to \Tcc\six)}{\hbox{Prob (cc}\to \Tcc\three)} \sim 1.5.
\end{eqnarray}
Similarly, we obtain the relations between $I=0$ and $I=1$ productions as
\begin{eqnarray}
&& \frac{\hbox{Prob (cc}\to \Tcc\three)^{I=1}_{J}}{\hbox{Prob (cc}\to \Tcc\three)} \sim 
\begin{cases}
\frac{1}{3}, & J=0 \\
1, & J=1 \\
\frac{5}{3}, & J=2
\end{cases} \\
&& \frac{\hbox{Prob (cc}\to \Tcc\six)^{I=1}}{\hbox{Prob (cc}\to \Tcc\six)} \sim 1
\end{eqnarray}
This simple model gives us a useful guidance on estimation of the production of $\Tcc$ accompanying light quark fragmentation.

\section{Numerical results}
\label{sec:numerical}

\subsection{Total cross sections}

\begin{figure}
\includegraphics[width=15cm]{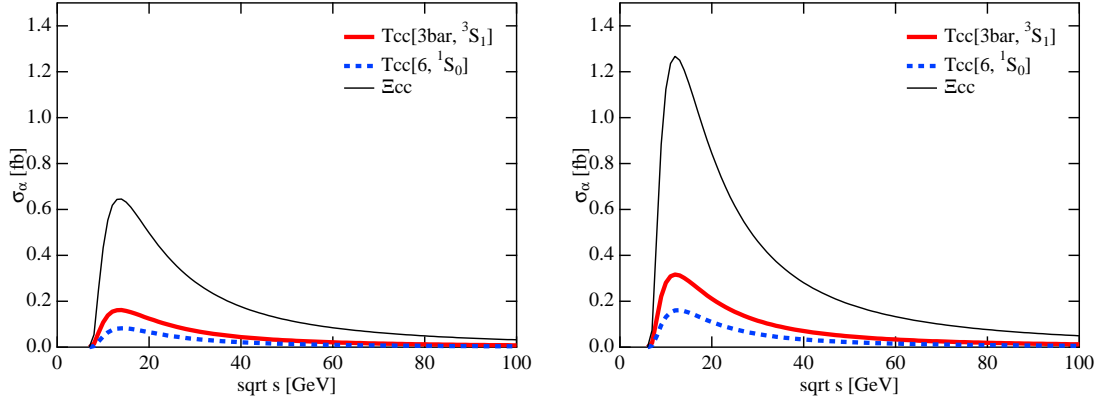}
\caption{Total cross sections of the reaction ${\rm e}^{+}{\rm e}^{-}\to \Tcc[\alpha]+{\rm X}$ with $\alpha=\three$ (Thick solid lines) and $\six$ (dashed lines). The thin solid lines represent the cross sections of the $\Xi_{\rm cc}$ production. The charm quark mass and the strong coupling constant are set to be $m_{\rm c}=1.8$ GeV and $\alpha_{s}=0.212$ ($m_{\rm c}=1.6$ GeV and $\alpha_{s}=0.221$) in the left (right) panel.
}
\label{fig:Totcross}
\end{figure}

Now we present the numerical results of the total cross section. In the left panel of Fig.~\ref{fig:Totcross}, we show the total cross sections $\sigma_{\alpha}$ of the $\Tcc$ production in the $e^{+}e^{-}$ collisions with $m_{\rm c}=1.8$ GeV for both color configurations. The cross sections of the $\Xi_{\rm cc}$ production are plotted for comparison. We note that the mass of $\Tcc$ is $2m_{\rm c}=3.6$ GeV in the present framework, while the estimation by the constituent quark model is about 3.8 GeV. The strong coupling constant is chosen as $\alpha_{s}=0.212$ at the scale $\mu_{R}=2m_{\rm c}$ \cite{Jiang:2012jt}, which was used for the production of $\Xi_{{\rm cc}}$ (see also Refs.~\cite{Berezhnoy:2003hz,Ma:2003zk} for early works on $\Xi_{{\rm cc}}$).
We show the cross sections when the values of $R_{cc}(0)$ in set A in table~\ref{table:R0} are used.
The values of the cross sections in model B are about 1/4 of those in model A.

%

We find that the cross section of $\alpha=\three$ is larger than that of $\six$. The difference is caused by the color-spin projection in Eqs.~\eqref{eq:projection3bar} and \eqref{eq:projection6}, as well as the probability factor in the light quark sector \eqref{eq:Prob3} and \eqref{eq:Prob6}.
In both cases, the cross sections start to increase at the threshold $2m_{c}$, as shown in the figures, and the peak of the production cross section is found around the energy slightly above the threshold ($\sqrt{s}\sim 10$ GeV).
Near the $Z$ boson pole ($\sqrt{s}\sim 90$ GeV), the cross section will be dominated by the contribution through the $Z$ production over the $\gamma^{*}$ production \cite{Jiang:2012jt}, but this is not included in the present discussion.
For larger $\sqrt{s}$, the cross section decreases as being proportional to $1/s$.

To estimate the dependence on the quark mass, we also calculate the cross section with $m_{\rm c}=1.6$ GeV. The strong coupling constant is scaled by the two-loop running formula
\begin{align}
    \alpha_{s}(\mu_{R})
    &=
    \frac{1}{b_{0}t}
    \Biggl(
    1-\frac{b_{1}}{b_{0}}\frac{\ln t}{t}
    \Biggr),
    \label{eq:scaling}
\end{align}
with $t=\ln(\mu_{R}^{2}/\Lambda^{2})$, $b_{0}=(33-2n_{f})/12\pi$, $b_{1}=(153-19n_{f})/24\pi^{2}$ and $n_{f}=3$. The nonperturbative scale $\Lambda = 0.239$ GeV is determined to reproduce $\alpha_{s}(3.6\text{ GeV})=0.212$. This leads to $\alpha_{s}(3.2\text{ GeV})=0.221$. In this case, the qualitative behavior of the cross section remains the same. The magnitude of the cross section increases by about factor two, mainly because of the increase of the phase space.

\begin{figure}
\includegraphics[width=15cm]{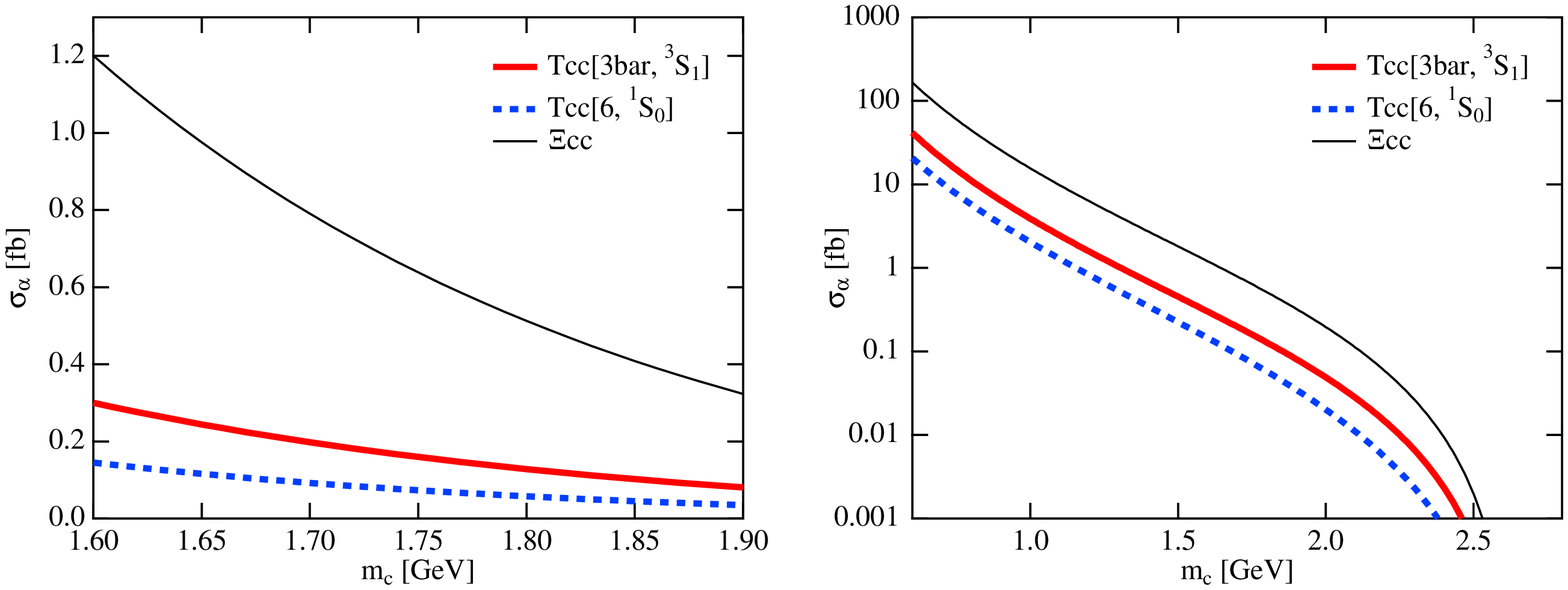}
\caption{The quark mass dependence on the total cross sections of the reaction ${\rm e}^{+}{\rm e}^{-}\to \Tcc[\alpha]+{\rm X}$ with $\alpha=\three$ (Thick solid lines) and $\six$ (dashed lines) at $\sqrt{s}=10.6$ GeV. The thin solid lines represent the cross sections of the $\Xi_{\rm cc}$ production. 
}
\label{fig:Totcross_mc}
\end{figure}

At the Belle energy $\sqrt{s}=10.6$ GeV, total cross section is estimated as
\begin{align}
    \sigma_{\alpha}
    =& 
    \begin{cases}
    0.0427 \ [{\rm fb}] & 
    \text{for } \alpha= \three^{I=1}_{J=0} \\
    0.1281 \ [{\rm fb}] & 
    \text{for } \alpha=\three, \three^{I=1}_{J=1} \\
    0.2135 \ [{\rm fb}] & 
    \text{for } \alpha= \three^{I=1}_{J=2} \\
    0.05762 \ [{\rm fb}] & 
    \text{for } \alpha=\six, \six^{I=1}
    \end{cases}
\end{align}
for $m_{\rm c}=1.8$ GeV and
\begin{align}
    \sigma_{\alpha}
    =& 
    \begin{cases}
    0.1002 \ [{\rm fb}] & 
    \text{for } \alpha=\three^{I=1}_{J=0} \\
    0.3007 \ [{\rm fb}] & 
    \text{for } \alpha=\three,\three^{I=1}_{J=1} \\
    0.5012 \ [{\rm fb}] & 
    \text{for } \alpha=\three^{I=1}_{J=2} \\
    0.1451 \ [{\rm fb}] & 
    \text{for } \alpha=\six, \six^{I=1}
    \end{cases}
\end{align} 
for $m_{\rm c}=1.6$ GeV. The quark mass dependence of the cross section around this region is shown in the left panel of Fig.~\ref{fig:Totcross_mc} where the strong coupling constant is also scaled according to Eq.~\eqref{eq:scaling}.
We obtain the similar values for $\Xi_{{\rm c}{\rm c}}$ in correspondence to $\Tcc\three$.
Those values are comparable with the upper limit for the $\Xi_{{\rm c}{\rm c}}$ cross section measured in Belle \cite{Kato:2013ynr}.
It may be interesting to compare the above numbers with the cross section of ${\rm e}^{+}{\rm e}^{-} \rightarrow {\rm J}/\psi\,{\rm c}\,\bar{\rm c}$ estimated in the NRQCD framework at leading order, 0.27 pb \cite{Zhang:2006ay}.\footnote{At the next-to-leading order, the authors in Ref.~\cite{Zhang:2006ay} obtained 0.47 pb, which is comparable with 0.74 pb measured in Belle \cite{Pakhlov:2009nj}.}
Our values given above for the $\Tcc + \bar{\rm c}\bar{\rm c}$ production are smaller than those for the ${\rm J}/\psi\,{\rm c}\,\bar{\rm c}$ production, as expected.

%

A naive extrapolation of the present result in much wider range of the quark mass is shown in the right panel of Fig.~\ref{fig:Totcross_mc}. The threshold of the reaction is $\sqrt{s}=4m_{\rm c}$, so the cross section vanishes at $m_{\rm c}=2.65$ GeV. When the quark mass is decreased, the cross section increases exponentially. At $m_{\rm c}=0.6$ GeV, the cross section is about a few pb order. This cross section, with multiplying the factor $1/4$ to account for the electric charge of the quarks, formally corresponds to the $\Xi$ production in the strangeness sector. Although this quark mass region is beyond the applicability of the NRQCD framework, we may regard it as a rough estimate of the $\Xi$ production via the hard process. The dominant production mechanism of the strangeness at $\sqrt{s}=10.6$ GeV is the color string breaking, so the contribution from the hard process should be smaller than the total cross section of the $\Xi$ production.
Nevertheless, it is interesting to observe that the present estimation may not be contradict with the experimental data measured by Belle, $\sim 26$ pb for $\Xi$ production in ${\rm e}^{+}{\rm e}^{-}$ collisions \cite{Sumihama:2014qfa}. 

\subsection{Differential cross sections}

Next we show the differential cross section with respect to the momentum and angle of the final $\Tcc$. To focus on the qualitative feature of the differential cross section, we calculate the normalized differential cross section defined as
\begin{align}
    \frac{1}{\sigma_{\alpha}}
    \frac{{\rm d}\sigma_{\alpha}}{{\rm d}p}
    =& 
    \frac{1}{\sigma_{\alpha}}
    \int \ {\rm d}\cos\Theta\frac{{\rm d}\sigma_{\alpha}}
    {{\rm d}p\ {\rm d}\cos\Theta},
    \label{eq:cross_section_theta}
\end{align}
and 
\begin{align}
    \frac{1}{\sigma_{\alpha}}
    \frac{{\rm d}\sigma_{\alpha}}{{\rm d}\cos\Theta}
    =& 
    \frac{1}{\sigma_{\alpha}}
    \int \ {\rm d}p \frac{{\rm d}\sigma_{\alpha}}
    {{\rm d}p\ {\rm d}\cos\Theta}.
    \label{eq:cross_section_p}
\end{align}
These quantities are useful, because they are independent of the nonperturbative matrix element $h_{\alpha}$.
Hence the results in this subsection are independent of the choice of set A or B in table~\ref{table:R0}.

\begin{figure}
\includegraphics[width=15cm]{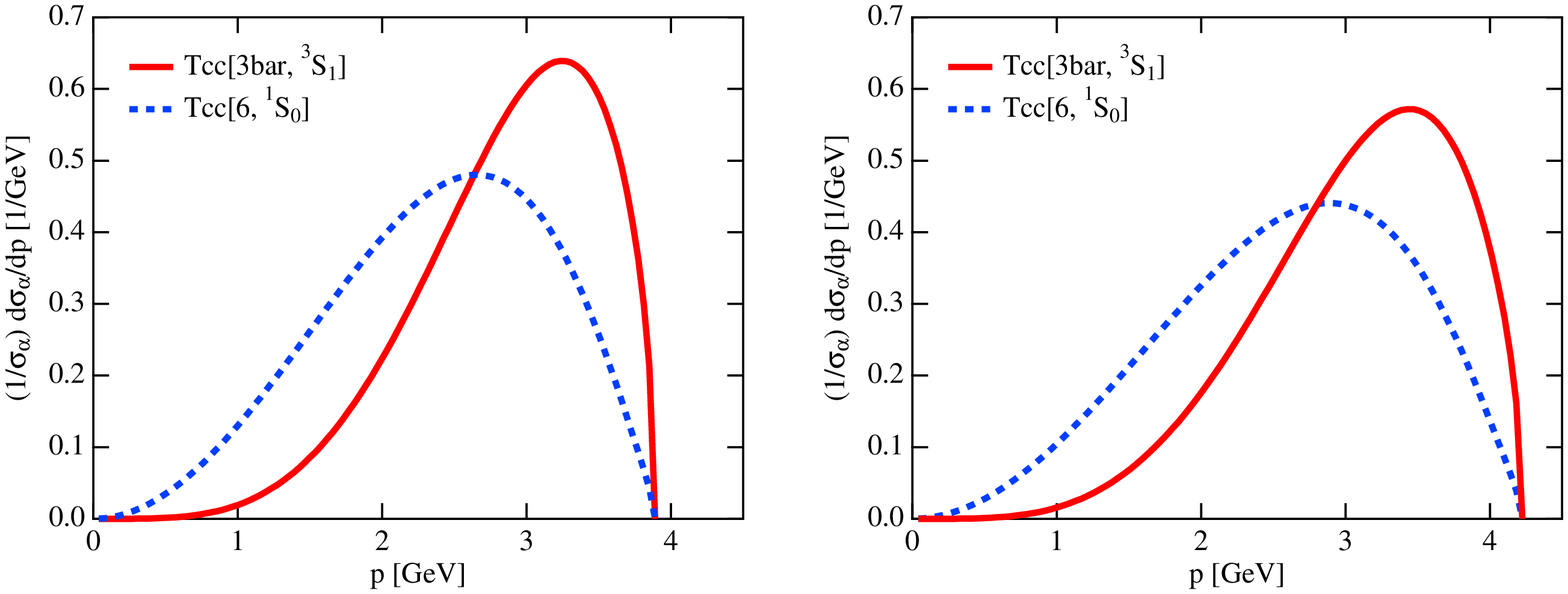}
\caption{Momentum dependence of the normalized differential cross sections $(1/\sigma_{\alpha}) {\rm d}\sigma_{\alpha}/{\rm d}p$ of the reaction ${\rm e}^{+}{\rm e}^{-}\to \Tcc[\alpha]+{\rm X}$ with $\alpha=[\bm{\bar{3}},{}^{3}{\rm S}_{1}]$ (solid lines) and $[\bm{6},{}^{1}{\rm S}_{0}]$ (dashed lines) at $\sqrt{s}=10.6$ GeV. The charm quark mass and the strong coupling constant are set to be $m_{\rm c}=1.8$ GeV and $\alpha_{s}=0.212$ ($m_{\rm c}=1.6$ GeV and $\alpha_{s}=0.221$) in the left (right) panel.
}
\label{fig:dcross_pvar}
\end{figure}

The results of the dependence on the momentum of $\Tcc$ at the Belle energy $\sqrt{s}=10.6$ GeV, Eq.~(\ref{eq:cross_section_theta}), are shown in Fig.~\ref{fig:dcross_pvar} with $m_{\rm c}=1.8$ GeV and 1.6 GeV. Irrespective of the quark mass, we find the following qualitative difference depending on the color configurations. The peak position of $\Tcc\three$ is concentrated in the large $p$ region. On the other hand, $\Tcc\six$ is produced even with a small momentum $p\sim 1$ GeV where only a tiny fraction of $\Tcc\three$ can be produced. 

%

\begin{figure}
\includegraphics[width=15cm]{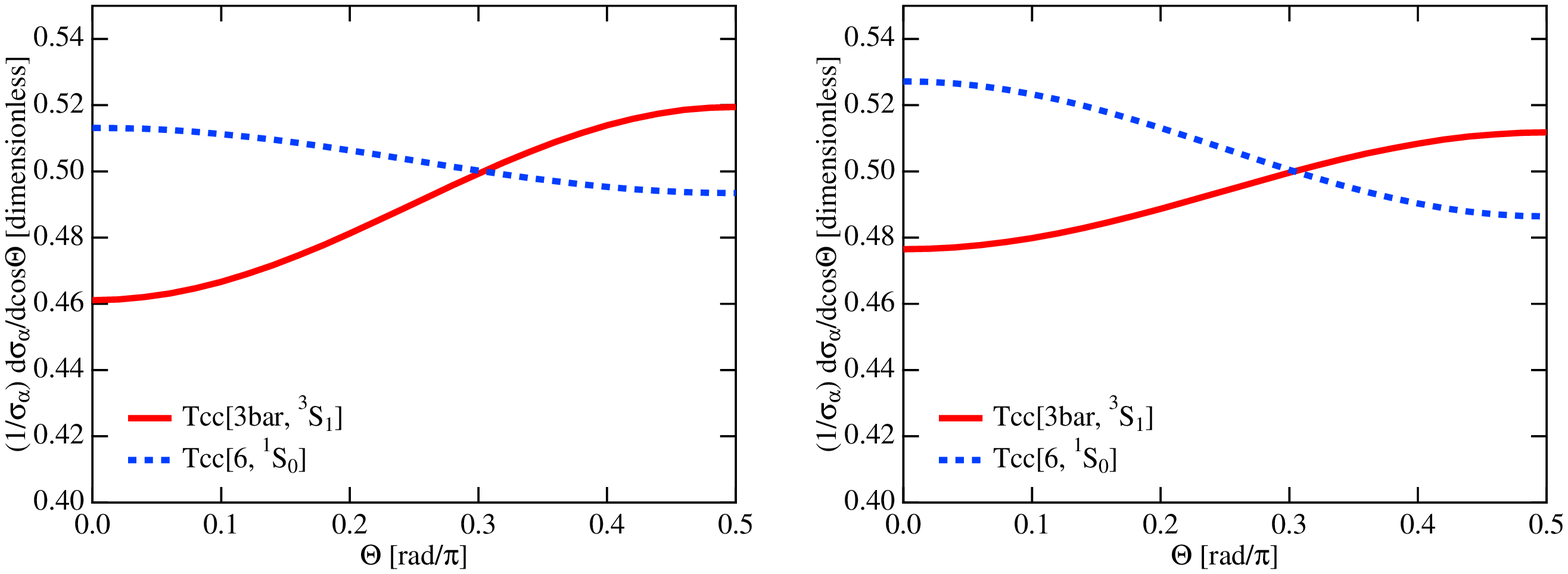}
\caption{Angular dependence of the normalized differential cross sections $(1/\sigma_{\alpha}) {\rm d}\sigma_{\alpha}/{\rm d}\Theta$ of the reaction ${\rm e}^{+}{\rm e}^{-}\to \Tcc[\alpha]+{\rm X}$ with $\alpha=\three$ (solid lines) and $\six$ (dashed lines) at $\sqrt{s}=10.6$ GeV. The charm quark mass and the strong coupling constant are set to be $m_{\rm c}=1.8$ GeV and $\alpha_{s}=0.212$ ($m_{\rm c}=1.6$ GeV and $\alpha_{s}=0.221$) in the left (right) panel.
}
\label{fig:dcross_Theta}
\end{figure}

Next we examine the dependence on $\Theta$ which is the production angle of $\Tcc$ measured from the collision axis. The results are shown in Fig.~\ref{fig:dcross_Theta}. The differential cross section has a symmetry under $\Theta\leftrightarrow \pi-\Theta$ because of the exchange of the electron and positron momenta. We find that $\Tcc\three$ is produced mainly in the transverse direction ($\Theta=\pi/2,3\pi/2$), while the peak of the $\Tcc\six$ is in the direction of the collision axis ($\Theta=0,\pi$).\footnote{If we decrease the quark mass further down to $\lesssim 1.4$ GeV, the peak of the $\Tcc\three$ production also lies in the direction of the collision axis. In this case, however, the mass of ${\rm T}_{\rm cc}$ is lower than 2.8 GeV (the binding energy from the $DD^{*}$ threshold is larger than 1 GeV). We consider such deep binding is unrealistic.} Although the magnitude of the angular dependence is not strong, the $\Tcc\three$ production has opposite angular dependence from that of the $\Tcc\six$.

%

In this way, we observe that the $\Tcc[\alpha]$ states with different internal color configuration $[\alpha]$ provide qualitatively different momentum/angular dependence of the cross section. This will help the experimental distinction of the internal color configurations of $\Tcc[\alpha]$.

\section{Summary}

We discuss the production of the doubly-charmed exotic hadron $\Tcc$ with quark content ${\rm c}{\rm c}\bar{\rm u}\bar{\rm d}$ from ${\rm e}^{+}{\rm e}^{-}$ collisions.
Based on the formalism of the NRQCD, we consider perturbative part of the productions of ${\rm c}{\rm c}\bar{\rm c}\bar{\rm c}$ via virtual gamma from ${\rm e}^{+}{\rm e}^{-}$ , and estimated the non-perturbative matrix elements from the non-relativistic quark model and the phenomenological estimation of the fragmentations of ${\rm c}{\rm c}$ quark pair.
The phenomenological discussion of the fragmentation, which was not considered in the previous work \cite{Hyodo:2012pm}, induces a suppression for the cross sections by some factors.
We also investigate the color structure of ${\rm c}{\rm c}$ quark pairs with different color-spin structure, $\Tcc\three$ and $\Tcc\six$.
As a result, we estimate the cross sections of $\Tcc$ as 
0.1281-0.3007 
fb for $\Tcc\three$ and 
0.05762-0.1451 
fb for $\Tcc\six$ depending on the parameter sets.
It is also interesting to observe that the momentum- and angle-dependence of the cross sections of $\Tcc\three$ are qualitatively different from those of $\Tcc\six$, and such information will be important to study the internal structure of $\Tcc$ in measurements in experiments.

\section*{Acknowledgments} 
%
S.Y. is supported by Grant-in-Aid for Scientific Research 
on Priority Areas ``Elucidation of New Hadrons with a Variety of Flavors 
(E01: 21105006)".
This work was partly supported by the Grant-in-Aid for Scientific Research from 
MEXT and JSPS (Grant Nos. JP16K17694, JP25247036, JP15K17641 and JP17K05435)
and by NNSFC (Grant No. 11275115).

\bibliographystyle{elsarticle-num} 

\end{document}